\documentclass[superscriptaddress,english,floatfix,twocolumn,showpacs,aps,prb,preprintnumbers, reprint]{revtex4-2}

\usepackage[normalem]{ulem}
\usepackage{amsmath}
\usepackage{bigints}
\usepackage{amsfonts}
\usepackage{amssymb}
\usepackage{dsfont}
\usepackage{bm}
\usepackage{adjustbox}
\usepackage{epstopdf}
\usepackage{color}
\usepackage{textcomp}
\usepackage{gensymb}
\usepackage{graphicx}
\usepackage[dvipsnames]{xcolor}

\usepackage{dcolumn}
\usepackage{bm}
\usepackage{braket}
\usepackage{hyperref}
\usepackage{booktabs}

\usepackage{hyperref}
\hypersetup{
    colorlinks=true,
    linkcolor=BurntOrange,
    urlcolor=BrickRed,
    citecolor=Mulberry,
    pdftitle={Adaptive variational ground state preparation for spin-1 models on qubit-based architectures},
    pdfauthor={Jo\~ao C. Getelina, Cai-Zhuang Wang, Thomas Iadecola, Yong-Xin Yao, Peter P. Orth}
}

\begin{document}

\title{Adaptive variational ground state preparation for spin-1 models on qubit-based architectures}

\begin{abstract}
We apply the adaptive variational quantum imaginary time evolution (AVQITE) method to prepare ground states of one-dimensional spin $S=1$ models. We compare different spin-to-qubit encodings (standard binary, Gray, unary, and multiplet) with regard to the performance and quantum resource cost of the algorithm. 
Using statevector simulations we study two well-known spin-1 models: the Blume-Capel model of transverse-field Ising spins with single-ion anisotropy, and the XXZ model with single-ion anisotropy. We consider system sizes of up to $20$ qubits, which corresponds to spin-$1$ chains up to length $10$. We determine the dependence of the number of CNOT gates in the AVQITE state preparation circuit on the encoding, the initial state, and the choice of operator pool in the adaptive method. Independent on the choice of encoding, we find that the CNOT gate count scales cubically with the number of spins for the Blume-Capel model and quartically for the anistropic XXZ model. However, the multiplet and Gray encodings present smaller prefactors in the scaling relations. These results provide useful insights for the implementation of AVQITE on quantum hardware.
\end{abstract}

\author{Jo\~{a}o C. Getelina}
\affiliation{Ames National Laboratory, Ames, Iowa 50011, USA}

\author{Cai-Zhuang Wang}
\affiliation{Ames National Laboratory, Ames, Iowa 50011, USA}
\affiliation{Department of Physics and Astronomy, Iowa State University, Ames, Iowa 50011, USA}

\author{Thomas Iadecola}
\affiliation{Ames National Laboratory, Ames, Iowa 50011, USA}
\affiliation{Department of Physics and Astronomy, Iowa State University, Ames, Iowa 50011, USA}

\author{Yong-Xin Yao}
\email{ykent@iastate.edu}
\affiliation{Ames National Laboratory, Ames, Iowa 50011, USA}
\affiliation{Department of Physics and Astronomy, Iowa State University, Ames, Iowa 50011, USA}

\author{Peter P. Orth}
\email{peter.orth@uni-saarland.de}
\affiliation{Ames National Laboratory, Ames, Iowa 50011, USA}
\affiliation{Department of Physics and Astronomy, Iowa State University, Ames, Iowa 50011, USA}
\affiliation{Department of Physics, Saarland University, 66123 Saarbr\"ucken, Germany}

\maketitle

\section{Introduction}
Recent advances in simulating quantum many-body systems on quantum processing units (QPUs)~\cite{asp_ipea,peruzzoVariationalEigenvalueSolver2014,
hardware_efficient_vqe,vqe_pea_h2,nisq,rmp_qcc,bauer2020quantum,cerezo2021variational, qite_chan20, Berthusen:2022, Chen2022,tilly2022variational,Huggins2022Unbiasingfq,kim2023scalable,vandenbergProbabilisticErrorCancellation2023,kimEvidenceUtilityQuantum2023}
have shown that some open questions in condensed matter physics may soon be addressed using quantum computations. Most suited are problems which involve quantum states whose entanglement grows with system size, which are difficult so simulate on classical computers. Examples are the preparation of ground and thermal states of nonintegrable many-body models in dimensions $d>1$ and the simulation of their nonequilibrium dynamics~\cite{setia2019,getelina2023adaptive,miessen2023quantum,shtanko2023uncovering, kimEvidenceUtilityQuantum2023}. 
One experimentally relevant class of models that provide a wealth of interesting open challenges are interacting spin models. While much work has focused on models with spin $S=1/2$, which allow for a direct identification of the spin with a qubit degree of freedom, many experimentally relevant materials host spins with $S> 1/2$. Since the Hilbert space dimension grows as $(2S+1)^L$ with the number of spins $L$, models with $S\geq 1$ require more memory than their $S=1/2$ counterparts and may thus especially benefit from a quantum computing approach. On the other hand, for qubit based hardware, they require an additional spin-to-qubit encoding step that can lead to increased qubit and gate-depth overhead. Although one may employ qudit platforms to avoid this step~\cite{ogunkoya2023qutrit,wang2023}, qubit-based hardware is currently the most readily accessible.

Physically, models with integer spin are known to behave quite differently from those with half-integer spins. For example, in one dimension it is well known that there exists a finite energy gap above the ground states of integer spin chains~\cite{haldane1983a,haldane1983b}, i.e., they exhibit a finite spin correlation length in the ground state, while half-integer chains are gapless, corresponding to an infinite correlation length~\cite{LIEB1961407}. The gapped phase of integer spin chains can be topologically nontrivial and exhibit fractionalized edge excitations with quantum numbers different from the bulk constituents~\cite{affleck1988,pollmann2010,polmann2012}. Models with spin $S \geq 1$ generally allow for terms that are absent in spin $S=1/2$ models. Prominent examples are single-ion anisotropy terms such as $(S^a)^2$ and biquadratic terms such as $(S^a_i S^b_j)^2$ ($a,b = x,y,z)$, which are often present in real materials. These terms not only contain important anisotropic contributions that can affect magnetically ordered states, but they can also include phenomena that cannot be realized in $S=1/2$ models such as quadrupolar (spin-nematic) order~\cite{papanicolaou1988,harada2002,broholm2005,arikawa2006,laeuchli2006}. Several known magnetic materials host spins with $S=1$ such as the triangular lattice materials NiGa$_2$S$_4$~\cite{broholm2005,nambu2006,shenoy2006} and Ba$_3$NiSb$_2$O$_9$~\cite{cheng2011high, quilliam2016gapless, faak2017evidence}, and the kagome spin-1 magnet YCa$_3$(VO)$_3$(BO$_3$)$_4$~\cite{miiller2011yca3}. 

Here, we focus on the task of determining the ground state phase diagram of spin $S=1$ models using a variational quantum algorithm (VQA) that can be implemented on qubit-based QPUs. We perform classical simulations of the algorithm to compare the qubit overhead and gate cost of four different binary spin-to-qubit encodings~\cite{sawaya2020,matteo2021}. VQAs are particularly suited for currently available noisy intermediate-scale quantum (NISQ) hardware~\cite{nisq}. Widely used variational algorithms include the variational quantum eigensolver (VQE)~\cite{peruzzoVariationalEigenvalueSolver2014, hardware_efficient_vqe,
vqe_pea_h2, vqe_theory, wecker2015_trotterizedsp, ho2019efficient, grimsleyAdaptiveVariationalAlgorithm2019,MayhallQubitAVQE, wiersema2020exploring, FengVQE}, subspace expansion methods~\cite{mcclean2017hybrid,mcclean2020,takeshita2020,qite_chan20,suchsland2021,yoshioka2022}, and methods based on quantum imaginary time evolution (QITE)~\cite{beach2019making,VQITE,qite_chan20,stokes2020quantum,qite_nla,smqite,
QITE_h2,Sun2021QuantumCO,Nishi2021ImplementationOQ}. Here, we employ the recently developed adaptive variational QITE (AVQITE) method~\cite{AVQITE} based on the variational QITE (VQITE) approach~\cite{VQITE}. Similar to adaptive versions of VQE~\cite{grimsleyAdaptiveVariationalAlgorithm2019, MayhallQubitAVQE, FengVQE,mukherjee2023comparative}, AVQITE adaptively builds a variational circuit ansatz during runtime. The ansatz circuit is designed so that the variational state follows the imaginary-time McLachlan equations of motion~\cite{VDynamics_Li}. We benchmark our classical simulations of the algorithm by comparing to exact diagonalization results and determine the scaling of the required quantum resources with the size of the simulated system. 

It is worth noting that VQITE with infinitesimal step sizes is equivalent to performing VQE with quantum natural gradient descent optimization~\cite{stokes2020quantum}. During AVQITE we closely monitor the McLachlan distance which measures the ability of the variational state to follow the exact imaginary time evolution path, and automatically augment the circuit ansatz if needed (i.e., if the McLachlan distance increases above a set threshold). Like in the original QITE algorithm~\cite{qite_chan20}, this procedure guarantees that AVQITE converges to the true ground state for large imaginary times $\tau$, provided we allow the circuit to grow to the required depth. In practice, we find that the final AVQITE ans\"atze are typically much shorter than those obtained from QITE~\cite{AVQITE}.

The remainder of the article is organized as follows. Models and methods are described in Sec.~\ref{sec:model}. Specifically, we introduce a class of spin-1 models in Sec.~\ref{sec:hamiltonians}, followed by a discussion of spin-1-to-qubit encodings in Sec.~\ref{sec:map}. The AVQITE algorithm is summarized in Sec.~\ref{sec:avqite}. We present the statevector simulation results in Sec.~\ref{sec:discussion}, with detailed descriptions of operator pool construction in Sec.~\ref{sec:pool}. We discuss the quantum circuit depth and its dependence on simulation hyperparameters in Sec.~\ref{sec:cnots}. Section~\ref{sec:discussion} gives a summary of results and an outlook.

\section{\label{sec:model}Models and methods}
Here, we present the family of model Hamiltonians that we investigate, which encompasses two well-known spin-1 models. We briefly review their physical behavior and phase diagrams. Then, we introduce in detail the four binary encodings used to map the spin-1 models onto a collection of qubits. Finally, we review the AVQITE method and define our benchmark criteria. 

\subsection{\label{sec:hamiltonians}General spin-1 chain Hamiltonian}
In this work we consider a one-dimensional spin-1 model that contains anisotropic XXZ-type nearest-neighbor exchange couplings, a biquadratic term describing single-ion uniaxial anisotropy, and an
external transverse magnetic field. The corresponding Hamiltonian reads
\begin{align}
	\mathcal{H} & = \sum_{j=1}^L J\left(S_{j}^{x}S_{j+1}^{x}+S_{j}^{y}S_{j+1}^{y}\right)+
	\Delta S_{j}^{z}S_{j+1}^{z} \nonumber \\
  & \qquad +D\left(S_{j}^{z}\right)^{2}+h_{x}S_{j}^{x} \,.
	\label{eq:hamiltonian}
\end{align}
Here, $S_{j}^{\alpha}$ ($\alpha = x,y,z$) are spin-1 operators at site $j$ that fulfill $[S_j^\alpha, S_k^\beta] = i \epsilon_{\alpha \beta \gamma} \delta_{jk} S_j^\gamma$ and $\sum_{\alpha} S_j^\alpha S_j^\alpha = S(S+1) = 2$. The parameters $J$ and $\Delta$ are the exchange coupling amplitudes in the $xy$-plane and along the $z$ direction, respectively, $D$ is the single-ion anisotropy strength, and $h_x$ is the magnetic field strength.

This Hamiltonian includes two well-studied models: first, the Blume-Capel (BC) model, which is obtained by setting $J=0$, and second, the XXZ model with single-ion anisotropy obtained by setting $h_x=0$. The BC model was initially devised to study the thermodynamics of mixtures in superfluid Helium~\cite{blume1966,capel1966,blume1971,kaufman1981}. 
Its two-dimensional classical counterpart is also known as the Blume-Emery-Griffiths model~\cite{berker1976,burkhardt1976}. The BC model represents the simplest generalization of the spin-1/2 transverse field Ising model
to higher spin. The single-ion biquadratic term $\propto (S_j^z)^2$ leads to an intricate phase diagram as a function of external field $h_x$ and single-ion anisotropy $D$, in which the ferromagnetic and paramagnetic phases are separated both by a continuous and first-order phase transition lines~\cite{alcaraz1985,qiu1986,balbao1987}. These two different phase transition types are joined at the so-called tricritical point, where the two ordered phases coexist with the disordered phase. In this work, we study the BC model in the vicinity of a continuous phase transition, where finding the ground state is more challenging than deep in the respective phases. Details on how to locate the phase transition from finite size studies are explained in Appendix~\ref{sec:ed}.

\begin{figure}
\begin{centering}
    \includegraphics[width=\linewidth]{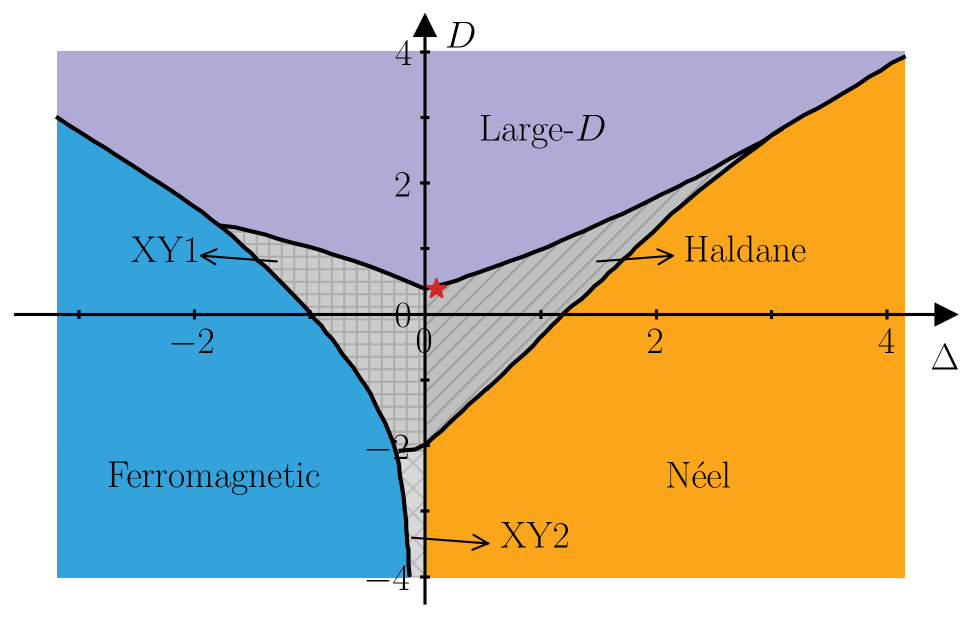}
    \caption{\label{fig:phase_diag}
	Reproduction of the phase diagram shown in Ref.~\cite{chen2003} of the spin-1 XXZ chain with single-ion anisotropy in the absence of external field, and with a fixed coupling amplitude $J=1$. The red star represents the parameters chosen in our subsequent investigation.}
\par\end{centering}
\end{figure}

The second model we study is the spin-1 XXZ model with a single-ion biquadratic term, which yields the phase diagram depicted in Fig.~\ref{fig:phase_diag}. Here the planar exchange coupling is set to $J=1$. There exist six different phases when tuning the strength of the exchange anisotropy $\Delta$ and the single-ion anisotropy $D$ while fixing an antiferromagnetic coupling $J = 1$. Here, we focus on the continuous phase transition between the Haldane and the large-$D$ phase, which is a topological phase transition. Details on our choice of parameters can also be found in Appendix~\ref{sec:ed}. 

\subsection{\label{sec:map} Spin-1-to-qubit encodings}
There are two options for simulating a system of quantum spins with $S > 1/2$ on a quantum computer. First, one can use a qudit-based quantum device (for a review see Ref.~\cite{wang2020}) where the number of qudit levels matches the number of spin states, $2S+1$. Alternatively, one can encode the spin-$S$ levels into qubits. Here, we focus on the second approach. It is worth noting that superconducting quantum computers based on transmon qubits also contain additional higher levels that can be accessed in principle~\cite{qiskit_qudit}. This has recently been proposed as an efficient way to simulate spin-$S$ models~\cite{ogunkoya2023qutrit,wang2023}.

One can write a single spin-1 degree of freedom in the computational basis of $n$ qubits with the aid of binary encodings~\cite{sawaya2020,matteo2021}. The authors of Ref.~\cite{sawaya2020} present a thorough study in which they estimate the quantum resources necessary for Trotterization in spin-$S$ systems. They have analyzed the results obtained for three different encodings: the standard binary encoding, the Gray code, and the unary encoding. Motivated by this approach, we also consider these three binary encodings when simulating quantum imaginary time evolution. In addition, we include a fourth encoding, which we refer to as the multiplet encoding. It is obtained from the angular momentum addition of two quantum spin-1/2 systems. The unary encoding uses $n=3$ qubits to represent a single spin-1 object, while the other three encodings use $n=2$ qubits. We refer to the total number of qubits used in the simulation as $N = nL$ in the following. 

In Table~\ref{tab:encoding} we show the correspondence between each of the
three spin-1 levels and the qubit computational basis states used to encode them. Each column corresponds to a different binary
encoding. Throughout this work we denote the spin-1 basis in bold, whereas non-bold numbers indicate qubit states.

\begin{center}
\begin{table}
\begin{centering}
\begin{tabular}{c|c|c|c|c}
 Spin-1 level& Standard & Gray & Unary & Multiplet \tabularnewline
\hline 
	$\left|\mathbf{0}\right\rangle $ & $\left|00\right\rangle $ & $\left|00\right\rangle $
		& $\left|001\right\rangle $  & $\left|00\right\rangle $\tabularnewline
	$\left|\mathbf{1}\right\rangle $ & $\left|01\right\rangle $ & $\left|01\right\rangle $
		& $\left|010\right\rangle $ & $\left(\left|01\right\rangle +\left|10\right\rangle
		\right)/\sqrt{2}$ \tabularnewline
	$\left|\mathbf{2}\right\rangle $  & $\left|10\right\rangle $ & $\left|11\right\rangle $
		& $\left|100\right\rangle $ & $\left|11\right\rangle $\tabularnewline
\hline 
\end{tabular}
\par\end{centering}
	\caption{\label{tab:encoding} The qubit computational basis representations of the three spin-1 levels (denoted in bold) for four different binary encodings.}
\end{table}
\par\end{center}

In addition to mapping the spin-1 levels at each site to qubit computational basis states, one needs to also transform the spin-1 operators into combinations of Pauli matrices. This operator mapping depends on the choice of encoding. For a given spin-1 operator $S^{\alpha}=\sum_{i,j=0}^{2}c^\alpha_{i,j}\left|\mathbf{i}\right\rangle \left\langle \mathbf{j}\right|$, one first has to apply the chosen binary encoding for the basis vectors $\left|\mathbf{i}\right\rangle $ and $\left|\mathbf{j}\right\rangle $. In the resulting expression, we can use the following set of equations~\cite{sawaya2020}
\begin{align}
\left|0\right\rangle \left\langle 0\right| & =\frac{1}{2}\left(\mathds{1}+\sigma^{z}\right),\label{eq:trafo1}\\
\left|1\right\rangle \left\langle 1\right| & =\frac{1}{2}\left(\mathds{1}-\sigma^{z}\right),\label{eq:trafo2}\\
\left|0\right\rangle \left\langle 1\right| & =\frac{1}{2}\left(\sigma^{x}+i\sigma^{y}\right),\label{eq:trafo3}\\
\left|1\right\rangle \left\langle 0\right| & =\frac{1}{2}\left(\sigma^{x}-i\sigma^{y}\right),\label{eq:trafo4}
\end{align}
where $\mathds{1}$ is the identity matrix and $\sigma^{x}$, $\sigma^{y}$, and $\sigma^{z}$ are the usual Pauli matrices. To provide an example, let us consider the spin-1 operator 
\begin{equation}
	S^{x}=\frac{1}{\sqrt{2}}\left[\left|\mathbf{0}\right\rangle \left\langle \mathbf{1}\right|+\left|\mathbf{1}\right\rangle
	\left\langle \mathbf{0}\right|+\left|\mathbf{1}\right\rangle \left\langle \mathbf{2}\right|+\left|\mathbf{2}\right\rangle 
	\left\langle \mathbf{1}\right|\right] \,.
\end{equation}
Choosing the Gray code as a binary encoding and mapping the basis according to Table~\ref{tab:encoding}, we find the qubit representation of the operator as
\begin{align}
	S^{x}\xrightarrow[\text{code}]{\text{Gray}}\widetilde{S^{x}}= \frac{1}{\sqrt{2}} & \left[\left|00\right\rangle
	\left\langle 01\right|+\left|01\right\rangle \left\langle 00\right| \right. \\\nonumber
	& \left. +\left|01\right\rangle \left\langle 11\right|+\left|11\right\rangle \left\langle 01\right|\right].
\end{align}
Finally, using Eqs.~\eqref{eq:trafo1}--\eqref{eq:trafo4} in each tensor factor of the above expression, we obtain
\begin{equation}
\widetilde{S^{x}}=\frac{\sqrt{2}}{4}\left(\mathds{1} \otimes \sigma^{x}+\sigma^{x}\otimes\mathds{1}+\sigma^{z}\otimes\sigma^{x}-\sigma^{x}\otimes\sigma^{z}\right).
\end{equation}
The same procedure is repeated for the other spin-1 operators within a chosen encoding, with the exception of the multiplet encoding. For the multiplet encoding, one should rather use the following expression in order to map spin-1 operators onto qubit operators:
\begin{equation}
	\widetilde{S^{\alpha}}=\frac{1}{2}\left(\sigma^{\alpha}\otimes\mathds{1}+\mathds{1}\otimes\sigma^{\alpha}\right),\label{eq:trafo-mp}
\end{equation}
with $\alpha=x,\,y,\,z$.

\subsection{\label{sec:avqite}Adaptive variational quantum imaginary time evolution}
Imaginary time evolution is a well-known approach to determine the ground state of a given model Hamiltonian $\mathcal{H}$. Starting from an initial state $\rho(0) = \ket{\psi(0)}\bra{\psi(0)}$, the density matrix evolves in imaginary time $\tau$ as 
\begin{equation}
\rho(\tau) = \frac{e^{-\mathcal{H} \tau} \rho(0) e^{-\mathcal{H}\tau}}{\text{Tr}[ e^{-2 \mathcal{H} \tau} \rho(0)]} \,.
\label{eq:rho_QITE}
\end{equation}
The denominator ensures a proper normalization of the density matrix. If the initial state has a finite overlap with the ground state of the model, the system  approaches the ground state for $\tau \rightarrow \infty$, as higher-energy states are projected out by the imaginary time propagator $e^{-\mathcal{H} \tau}$. In practice, however, this method suffers from the fact that exponentiating a matrix with unknown eigenvalues is numerically costly. One way to proceed is by noticing that after encoding the model Hamiltonian onto a collection of qubits, it consists of a linear combination of Pauli strings: $\mathcal{H} = \sum_j a_j P_j$. Assuming a small imaginary time step $\Delta\tau$, we can then apply a Trotter decomposition of $e^{-\tau\mathcal{H}}$ into a product of Pauli-string exponentials that can be more easily evaluated numerically. This procedure, however, is still limited by the memory of a classical computer, as the matrix dimension grows exponentially with system size, suggesting a potential advantage for quantum computing. 

The original QITE algorithm proposed in Ref.~\cite{qite_chan20} expressed a single imaginary time step $\tau \rightarrow \tau + \Delta \tau$ (including the renormalization of $\rho$) as a unitary operation that is obtained by solving a least squares problem. The classical complexity of this problem and the required number of measurements are exponentially large in the physical correlation length $\xi$ of the simulated system. On the other hand, the cost is only polynomial in the system size $N$ in space and time, an exponential savings compared to a purely classical implementation. While the algorithm is guaranteed to converge to the ground state, the depth of the unitary increases over time as $e^{\xi^d \tau}$, where $d$ is the dimensionality of the system and $\tau = N_\tau \Delta \tau$ with $N_\tau$ being the number of Trotter steps. This makes the approach infeasible on current NISQ hardware already for moderate values of $N_\tau$.

VQITE is an alternative method~\cite{VQITE,theory_vqs,stokes2020quantum} that works with a fixed variational ansatz given by $\ket{\psi(\boldsymbol{\theta})} = U(\boldsymbol{\theta}) \ket{\psi(0)}$. The variational parameters $\bm\theta$ in the parameterized quantum circuit $U(\bm \theta)$ obey a differential equation that guarantees that $\ket{\psi[\boldsymbol{\theta}(\tau)]}$ follows as closely as possible the imaginary time evolution defined by 
\begin{equation}
\frac{\partial \rho}{\partial \tau} = - \{\mathcal{H}, \rho(\tau)\} + 2 \text{Tr}[\mathcal{H} \rho(\tau)] \rho(\tau) \,.
\label{eq:QITE_diff_eq}
\end{equation}
Specifically, one minimizes the McLachlan distance
\begin{equation}\label{eq:mclachlan}
  \mathcal{L}^2 = \left\|\sum_{i = 1}^{N_\theta}\frac{\partial \rho[\boldsymbol{\theta}(\tau)]}{\partial \theta_i} \dot{\theta}_i +  \{\mathcal{H}, \rho[\boldsymbol{\theta}(\tau)]\} - 2 \langle\mathcal{H} \rangle_{\rho[\boldsymbol{\theta}(\tau)]} \rho[\boldsymbol{\theta}(\tau)] \right\|^2
\end{equation}
where $\| \mathcal{O} \| = \text{Tr}[ \mathcal{O}^\dag \mathcal{O}]$, $N_\theta$ is the total number of variational parameters, and we have used Eq.~\eqref{eq:QITE_diff_eq}. Straightforward manipulations yield that $\mathcal{L}^2$ can be expressed in terms of the (real part of the) quantum geometric tensor 
\begin{equation}
G_{ij}[\boldsymbol{\theta}(\tau)] = \left\langle \frac{\partial \psi}{\partial \theta_i}\Big|\frac{\partial \psi}{\partial \theta_j}\right\rangle + \left\langle\psi\Big|\frac{\partial \psi}{\partial \theta_i}\right\rangle \left\langle\psi\Big|\frac{\partial \psi}{\partial \theta_j}\right\rangle\,,
\label{eq:QGT_Re}
\end{equation}
the gradient 
\begin{align}
    V_i = - \frac12 \frac{\partial}{\partial \theta_i} \braket{\psi|\mathcal{H}|\psi} = - \text{Re} \left \langle \frac{\partial \psi}{\partial \theta_i}| \mathcal{H}|\psi\right \rangle,
\end{align}
and the variance of $\mathcal{H}$ in the state $\rho[\boldsymbol{\theta}(\tau)]$ as
\begin{equation}
L^2 = 2 \sum_{i,j} g_{ij} \dot{\theta}_i \dot{\theta}_j - 4 \sum_i V_i \dot{\theta}_i + 2 \left( \langle \mathcal{H}^2 \rangle - \langle \mathcal{H} \rangle^2 \right) \,,
\label{eq:McLachlan_final}
\end{equation}
where $g_{ij}=\text{Re}[G_{ij}] $ is the Fubini-Study metric. These quantities can be measured on quantum hardware~\cite{VQITE,theory_vqs,AVQITE}. Here, however, we compute $g_{ij}$, $V_i$ and $(\langle \mathcal{H}^2 \rangle - \langle \mathcal{H} \rangle^2)$ on a classical computer, since the main goal of this study is to perform benchmarking calculations and compare the properties of different encodings. 
Minimizing $\mathcal{L}^2$ using the variational principle $\delta \mathcal{L}^2= 0$ yields the equations of motion for the variational parameters
\begin{equation}
\sum_{j=1}^{N_\theta} g_{ij} \dot{\theta}_j = V_i \,.
\label{eq:EOM_theta}
\end{equation}
This form of the equations shows that VQITE with infinitesimal step size $\delta \tau$ corresponds to performing VQE with a quantum natural gradient descent optimizer, where we update the variational parameters according to $\sum_j g_{ij} \Delta \theta_j = -\frac{\partial}{\partial \theta_i} C(\boldsymbol{\theta}) \Delta \tau = V_i \Delta \tau$ with cost function $C(\boldsymbol{\theta}) = \frac12 \braket{\psi|\mathcal{H}|\psi}$~\cite{stokes2020quantum}. 

One of the main drawbacks of VQITE is that the algorithm is no longer guaranteed to converge to the true ground state of the system, since the state is confined to the variational manifold set by the parametrized circuit ansatz.
One way to resolve this issue is to let the ansatz expand dynamically and adaptively during runtime. In the AVQITE algorithm~\cite{AVQITE} the circuit expansion is controlled by monitoring the McLachlan distance $\mathcal{L}^2$ during time evolution. Whenever $\mathcal{L}^2$ is found to be above a user-defined threshold  in a given timestep (here we use $10^{-2}$), the circuit is enlarged by a unitary operator whose generator is chosen from a predefined operator pool. We discuss different choices of operator pools for our problem in the following section. The adaptive algorithm then selects the generator that leads to the maximal decrease of $\mathcal{L}^2$ and adds the corresponding unitary to the ansatz. The imaginary time evolution proceeds until the maximum energy gradient $\max_i |V_i|$ is less than a fixed cutoff, which we set to  $10^{-4}$ here. As a technical note, we mention that the Fubini-Study metric tensor $g_{ij}$ can have a large condition number, which we deal with by using Tikhonov regularization $g \rightarrow g + \lambda \openone$ with $\lambda$ the Tikhonov regularization parameter. We find that setting $\lambda=10^{-6}$ works well for the systems studied here.

In order to verify whether the final converged AVQITE state $\left|\Psi_{\text{AV}}\right\rangle$ is a good approximation of the true ground state of the system, we compute the fidelity 
\begin{equation}
\mathcal{F}=\left|\left\langle \Psi_{\text{AV}}\right.\left|\Psi_{\text{ED}}\right\rangle \right|^{2} \,.
\label{eq:fid}
\end{equation}
Here, $\left|\Psi_{\text{ED}}\right\rangle$ is the exact ground state found via exact diagonalization. We consider a  simulation to be successful if $\mathcal{F}\geq99.9\%$.

Finally, even though the calculations here are performed on classical computers, one can still estimate what would be the equivalent cost on quantum hardware. The main source of error on NISQ devices stems from the entangling gates such as CNOT. Hence, we use the final CNOT gate count as a figure of merit for the quantum hardware cost. For simplicity, we assume all-to-all connectivity of the qubits as is realized on trapped-ion based quantum hardware architectures. In this case, the number of CNOTs required to evaluate a Pauli string with $N_p$ Pauli operators is simply given by
\begin{equation}\label{eq:cnot}
	N_{CX} = 2\left(N_p - 1\right)\,.
\end{equation}

\section{\label{sec:results}Statevector simulations}
In this section we present our numerical results on ground-state properties of spin-1 models using the AVQITE method. We first discuss our choice of operator pool, and then we provide an estimate of the quantum resources required for accurate ground state calculations. All our results are obtained using noiseless statevector simulations on classical computers.

\subsection{\label{sec:pool}Operator pool construction}

As detailed in Sec.~\ref{sec:avqite}, the parameterized AVQITE state is constructed by applying unitary operators with generators drawn from a predetermined pool to a given initial reference state $\ket{\psi(0)}$. One can construct such operator pool by considering any combination of Pauli operators acting on each system site, provided it contains an odd number of $\sigma^{y}$ operators in order to ensure a real exponent for the imaginary time evolution.
However, as the number of possible combinations of Pauli operators increases exponentially with system size, we select a subset of these operators to construct the pool.

We consider two different choices of operator pool. First, we use a ``minimal pool'' that contains all possible permutations of the set 
\begin{equation}
\left\{ \sigma_{i}^{y},\sigma_{i}^{y}\sigma_{j}^{z}\right\}
\label{eq:pool_min}
\end{equation}
for $i\ne j\in \left[1, N = nL\right]$. For example, the minimal pool for a system with two qubits is given by the set  $\left\{\sigma_{1}^{y},\,\sigma_{2}^{y},\, \sigma_{1}^{y}\sigma_{2}^{z},\,\sigma_{1}^{z}\sigma_{2}^{y}\right\}$. The minimal pool has proven to be successful for the spin-1/2 models studied in Ref.~\cite{AVQITE}. In addition to the minimal pool, in this work we consider the following ``maximal pool'' 
\begin{equation}
\left\{ \sigma_{i}^{y},\,\sigma_{i}^{y}\sigma_{j}^{z},\, \sigma_{i}^{y}\sigma_{j}^{x},\,\sigma_{i}^{y}\sigma_{j}^{x}\sigma_{k}^{z}\right\}
\label{eq:pool_max}
\end{equation}
for $i\ne j\ne k \in \left[1, N=nL\right]$. The consideration of two different operator pools has been motivated by preliminary results which showed that the minimal pool does not exhibit good convergence properties for certain choices of reference state $\ket{\psi(0)}$ and binary encoding for the spin-1 models studied here.

It is worth emphasizing that these pools allow the ansatz state to explore qubit states that lie outside the spin-1 subspace. We choose these general qubit pools since we found operator pools based on spin-1 operators, for which the ansatz state remains in the spin-1 subspace (if the reference state lies in that subspace), to yield unsatisfactory results. For example, when simulating the BC model for chains with lengths $L=2,\,3$, using the pool $\{S^y_i, S^y_i S^z_i\}$ ($i=1,\ldots, L$) made of spin-1 operators, we were unable to reach a fidelity $\mathcal{F}$ of 99.9\% [see Eq.~\eqref{eq:fid}] for any choice of reference state in the qubit computational basis and for any of the four binary encodings. Moreover, even for the cases in which we have observed a fidelity between 
99\%-99.9\%, the number of required entangling gates was found to be much larger than for the qubit based pools defined in Eqs.~\eqref{eq:pool_min} and~\eqref{eq:pool_max}.

To monitor whether the final AVQITE state lies in the spin-1 subspace when using the qubit based pools, we introduce the local operators 
\begin{equation}
P_j=\frac{1}{2}\bigl[ (S^x_j)^2 + (S^y_j)^2 + (S^z_j)^2] \,,
\label{eq:P_j}
\end{equation}
where $S^\alpha_j$ are components of the spin-1 angular momentum operators. The $P_j$ operators act as the identity operator on spin-1 degrees of freedom but, when represented in terms of qubit operators via  Eqs.~\eqref{eq:trafo1}--\eqref{eq:trafo4} and Eq.~\eqref{eq:trafo-mp}, become projectors onto the spin-1 subspace of the qubit Hilbert space on site $j$. We define the global projection operator into the spin-1 subspace as $P = \bigotimes_{j=1}^L P_j$. For any state that belongs to the spin-1 Hilbert space, one obtains the expectation value  $\left\langle P \right\rangle = 1$. Hence, by measuring the global operator $P$ during imaginary time evolution, we monitor whether the final AVQITE state lies within the spin-1 subspace as desired.

\begin{figure*}[!ht]
\begin{centering}
	\includegraphics[width=\textwidth]{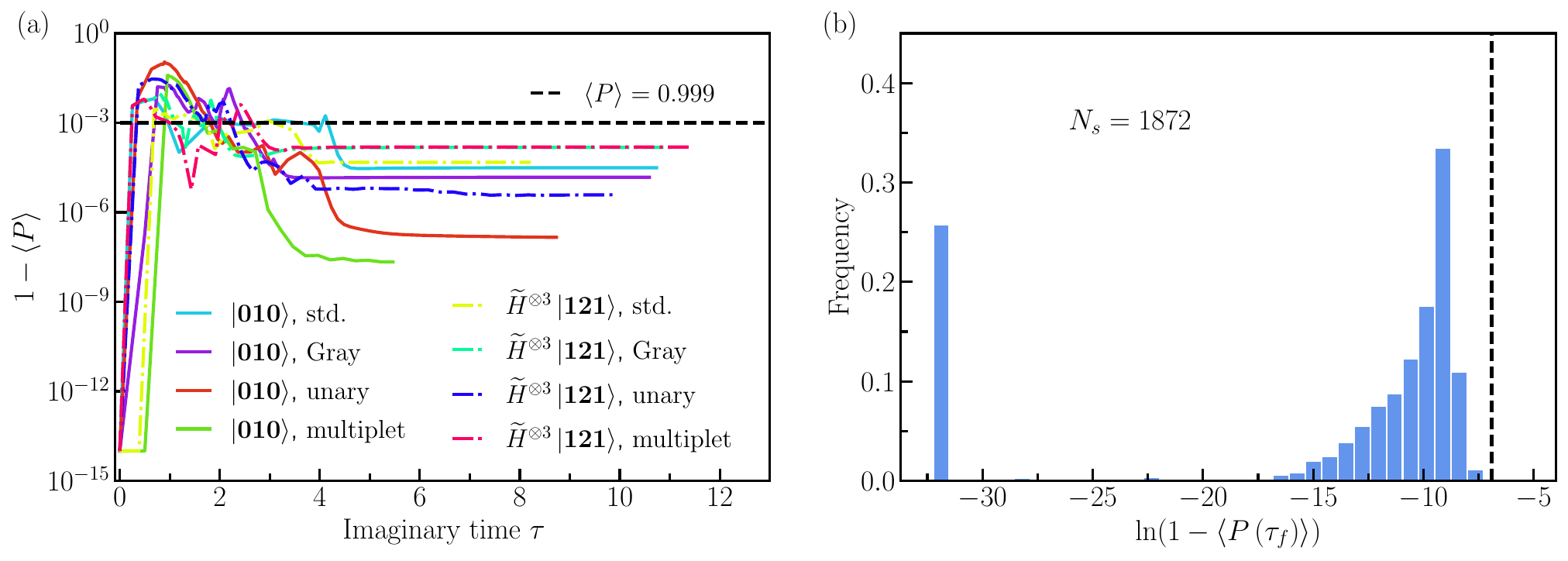}
  \caption{\label{fig:proj}
  (a) Expectation value of the spin-1 projection operator $\langle P \rangle= \left\langle \bigotimes_{j=1}^L P_j \right\rangle$, with $P_j$ defined in Eq.~\eqref{eq:P_j}, for the spin-1 BC model. We set $L=3$ and use the parameters $D/|\Delta| = -0.1$, $h_x/|\Delta| =-1.405$, and $\Delta = -1$. Panel (a) shows the time evolution of $\left\langle P \right\rangle$ under AVQITE for two different initial reference states and various binary encodings as shown in the legend. We use the maximal operator pool in all cases. The operator $\widetilde{H}$ corresponds to the Hadamard transform on the qubits in the corresponding encoding.
	(b) Distribution of $\langle P(\tau) \rangle$ at the final step of the simulation, $\tau = \tau_f$. Results are for a total of $1872$ different reference states and for system sizes $L=2,\,3,$ and $4$. In both panels the dashed line denotes the threshold $\left\langle P \right\rangle = 0.999$.}
\par\end{centering}
\end{figure*}

In Fig.~\ref{fig:proj} we show results for the expectation value of the projection operator $P$ for the spin-1 BC model with $L=3$. We set the parameters to be near a continuous phase transition: $J=0,\,\Delta=-1,\,D=-0.1,\,h_{x}=-1.405$ in Eq.~\eqref{eq:hamiltonian}. Panel~\hyperlink{fig:proj}{(a)} shows $\langle P(\tau) \rangle$ along the imaginary time evolution path for two different initial states and for all four binary encodings. For each simulation we have used the maximal pool, Eq.~\eqref{eq:pool_max}. We observe that every curve slightly drifts out of the spin-1 sector after the initial steps of the procedure (the initial value of $10^{-14}$ corresponds to zero within machine precision). However, as time progresses, the curves decrease again and fall below our chosen threshold of $\left\langle P\right\rangle =0.999$ [dashed line in Fig.~\ref{fig:proj}(a)].

\begin{figure*}
\begin{centering}
\includegraphics[width=\textwidth]{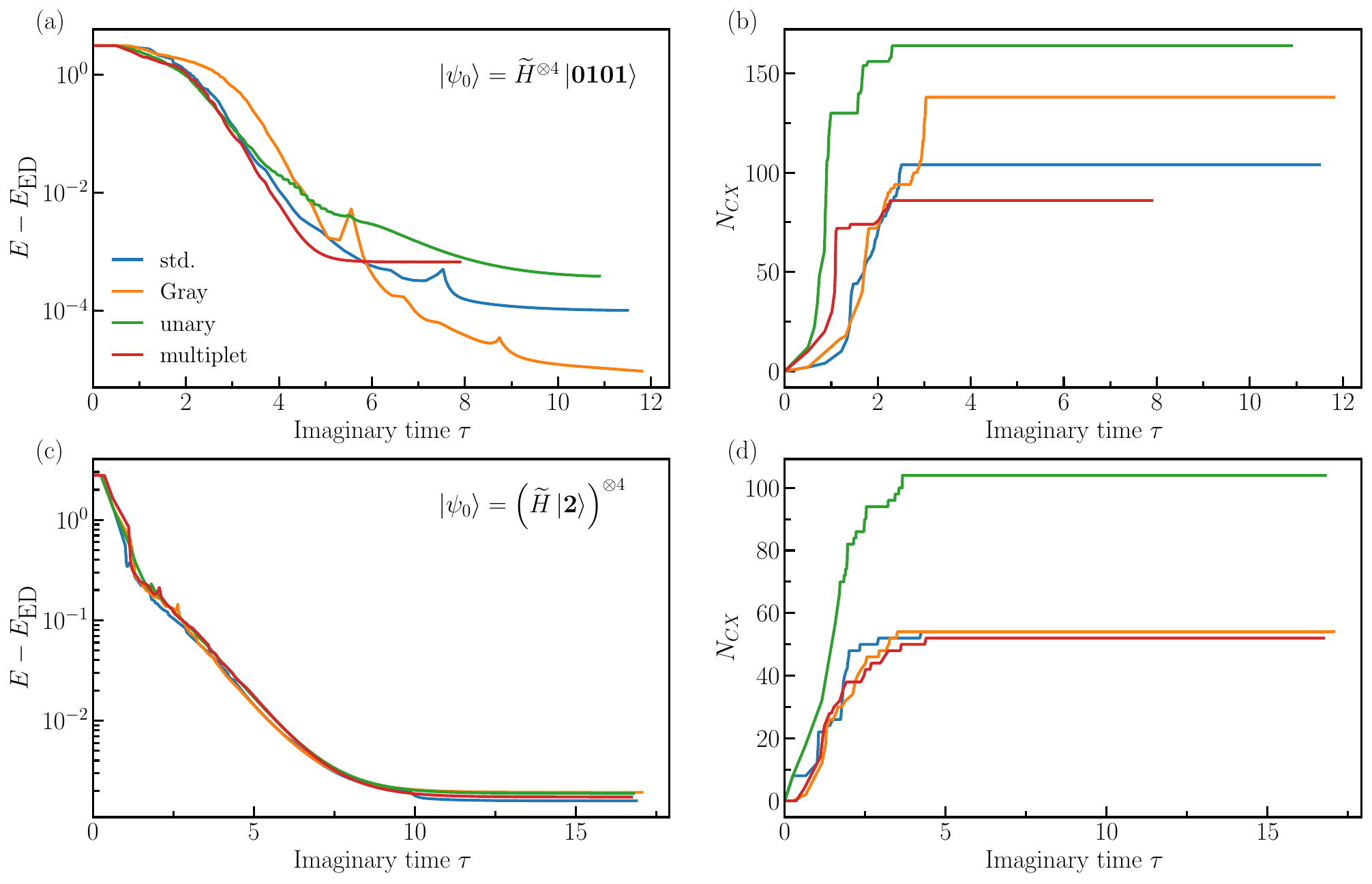}
  \caption{\label{fig:E-v-t}Imaginary-time evolution of the energy $E$ (a,c) and number of CNOT gates
  $N_{CX}$ (b,d) for the AVQITE simulation of the spin-1 XXZ model with single-ion anisotropy
  near criticality (for Hamiltonian parameters see text). Each panel shows data for each of the four studied binary encodings,
  Top and bottom panels use different reference states $\left|\psi_0 \right\rangle$, as specified in each plot.}
\par\end{centering}
\end{figure*}

In Fig.~\ref{fig:proj}\hyperlink{fig:proj}{(b)}, we show a histogram of the expectation value of $P$ at the final AVQITE step. We group results for chains with $L=2,\,3,\,4$ spin-1 sites, and for all possible product state reference states in the $x$ and the $z$ bases of the qubits. We include results for each of the four encodings and for both the minimal and the maximal pool. The histogram thus contains a total of $N_s=2\times2\times4\times\left(3^2+3^3+3^4\right)=1872$ simulations. We observe that all simulations lie within the spin-1 sector and have a weight outside that sector smaller than $10^{-3}$ (dashed line). In addition, from this sample we have verified an 83\% success rate to converge to the ground state, taking the fidelity $\mathcal{F}\ge0.999$ as the convergence criterion. This validates our choice of the qubit pools \eqref{eq:pool_min} and \eqref{eq:pool_max} and suggests that, once a pool is selected, convergence to a final state in the relevant spin-1 sector is likely to be achieved without trying a large number of initial reference states.  

When comparing the different encodings, we find that the standard, Gray and unary encodings all show about 10\% of samples reaching a final state with $\ln \left(1 - \left\langle P(\tau_f) \right\rangle\right)\approx 10^{-14}$ (leftmost histogram bin). For the multiplet encoding only about 2.5\% of final states reach that value, i.e., the multiplet encoding produces final states with larger weight outside the spin-1 subspace. We discuss the dependence of the convergence and the depth of the final AVQITE circuit on these hyperparameters in more detail in the next section.

\subsection{\label{sec:cnots}Quantum circuit depth for different simulation hyperparameters}
In this section we discuss in detail how the AVQITE circuit depth depends on the different hyperparameters of the algorithm: (1) encoding, (2) operator pool, and (3) reference state $\ket{\psi(0)}$. We have already established in the last section that the general qubit pools defined in Eqs.~\eqref{eq:pool_min} and~\eqref{eq:pool_max} work well for the models we consider. Note that while the convergence of the AVQITE method to the true ground state is guaranteed provided the reference state $\ket{\psi(0)}$ has finite overlap with the ground state (and exhibits nonzero initial energy gradient $V_i$), the dependence of the final circuit depth on $\ket{\psi(0)}$ is not known \emph{a priori}. 

To study the hyperparameter dependence in more detail, we first consider the XXZ model with single-ion anisotropy near the quantum phase transition between the ``large-$D$'' and the Haldane phase. We set the model parameters in Eq.~\eqref{eq:hamiltonian} to $J=1,\;\Delta=0.1,$ and $D=0.385$. In Fig.~\ref{fig:E-v-t}, we show how the energy and the number of CNOT gates $N_{CX}$ [as defined in Eq.~\eqref{eq:cnot}] evolve during imaginary time evolution as the AVQITE quantum circuit is expanded. The algorithm terminates whenever the maximum energy gradient reaches the threshold of $\text{max}_j |V_j| < 10^{-4}$. We compare the behavior of the four different encodings for two different initial states: in the upper panels of the figure, we consider the initial state $\ket{\psi(0)} \equiv \ket{\psi_0} = \widetilde{H}^{\otimes 4} \left| \mathbf{0101} \right\rangle$, where $\widetilde{H}$ is a Hadamard gate mapped to the corresponding binary encoding,  whereas in the bottom panels $\ket{\psi(0)} \equiv \ket{\psi_0} = \left| \mathbf{2} \right\rangle^{\otimes 4}$. The ansatz operators are drawn from the maximal pool [see Eq.~\eqref{eq:pool_max}]. While we observe a strong dependence of the energy convergence on the choice of encoding for the reference state in panel (a) (with all cases converging to the final energy with $|E - E_{\text{ED}}| < 10^{-3}$), we do not observe any dependence on encoding for the reference state of panel (c). Regarding the CNOT gate count shown in Fig.~\ref{fig:E-v-t}\hyperlink{fig:E-v-t}{(b,~d)}, we find that the unary encoding yields the largest gate count, which is a consistent observation throughout our study.

We now systematically study the dependence of the gate depth on the reference state. For this, we perform simulations of both models using all possible qubit $x$- and $z$-basis product states as the reference states. We consider both maximal and minimal operator pools, all four encodings, and set the system size to $L=4$ spin-1 sites, keeping the model parameters the same as those given in the beginning of this section. As a figure of merit, we use the CNOT gate count of the final AVQITE circuit at $\tau = \tau_f$, which we denote as $N_{CX}(\text{final step})$. The total number of simulations for each model is thus given by $16\times3^4 = 1296$. 

\begin{figure*}
\begin{centering}
\includegraphics[width=\textwidth]{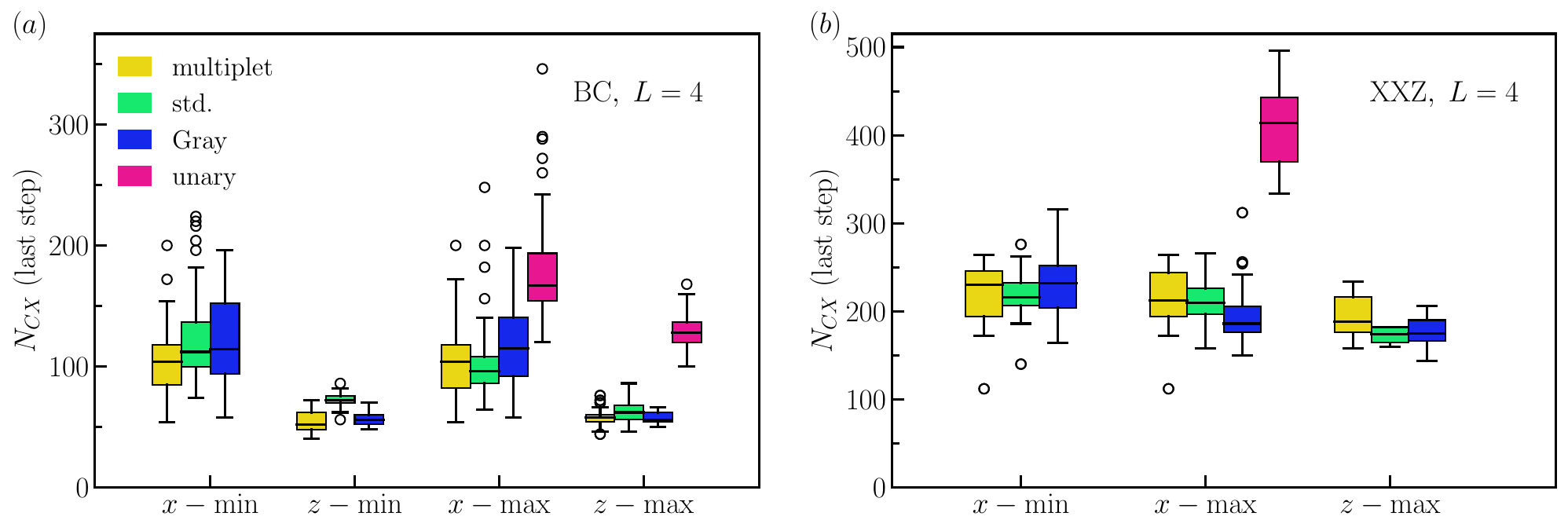}\caption{
  \label{fig:box}Box plots showing the distribution of number of CNOTs $N_{CX}$ at $\tau = \tau_f$ for (a) BC model, and (b) XXZ chain with parameters described in the main text. For each model we consider standard, Gray, multiplet and unary encodings. The corresponding  box distribution is drawn from simulating all possible 81 bitstrings in a system of size $L=4$. We discard samples where the final fidelity with the exact state wa less than $\mathcal{F}<0.999$. Each box encloses the interval between the first and third quartiles of the sample population, while the whiskers stretch out to $1.5$ times the interquartile length. Outliers are shown as open circles. Each horizontal-axis tick denotes the 
  corresponding product basis direction ($x$ or $z$) of the reference state and the operator pool choice (minimal or maximal)}
\par\end{centering}
\end{figure*}

The results are summarized in the box plots in Fig.~\ref{fig:box}. The plots display the distribution of $N_{CX}(\text{last step})$ as a colored bar bounded by the $25$th and $75$th percentiles with the horizontal line denoting the median value. The bars also contain vertical whiskers that encompass the data within $1.5$ times the interquartile length. Data outliers beyond this range are represented as open circles. The horizontal axis labels in Fig.~\ref{fig:box} corresponds to the $x$ or $z$ basis product initial states and using the minimal (min) or maximal (max) pool. The two panels
of Fig.~\ref{fig:box} refer to the two models we study. We exclude runs where the final fidelity with the exact ground state was smaller than 99.9\% and we also exclude reference states with vanishing initial energy gradient for which the algorithm halts at the first step. With these criteria we observe a success rate of 82\% for the BC model and 64\% for the XXZ model. 

For the BC model we find that the unary encoding yields by far the largest $N_{CX}$. The three other encodings yield comparable $N_{CX}$ with the multiplet encoding having a slightly smaller median $N_{CX}$ (except for the $x$-max configuration). The minimal gate count of about $N_{CX} \approx 50$ is observed for $z$-min. The variance is larger for qubit $x$-basis initial states than for the $z$-basis states. Turning to the XXZ model, we observe that that $N_{CX}$ is generally about a factor of two larger. Again, the unary encoding yields by far the largest $N_{CX}$ and the three other encodings produce similar $N_{CX}$. The smallest gate count median of $N_{CX} \approx 200$ is found for the standard and Gray encodings in the $z$-max configuration. Note that there was not a single successful run for the XXZ model using the minimal pool and qubit $z$-basis initial states and we do not show unary results when the gate count was too high and/or not sufficiently many runs converged. 

In order to find a direct relation between the reference state conditions and $N_{CX}(\text{last step})$, we ordered our data according to the following values (from most to least important): the final $N_{CX}$, the total $N_{CX}$ count used during imaginary time evolution, the number of imaginary time steps needed to reach the energy gradient threshold, and the pool choice. With these criteria we were able to identify that the best results in terms of final CNOT gate count all share a similar energy variance in the initial reference state. However, since there are many initial states that exhibit a similar energy variance (some of which lead to a larger than average $N_{CX})$, this criterion does not uniquely identify the best reference states (that require small $N_{CX}$).  

Next, we establish a scaling relation between $N_{CX}(\text{last step})$ and the system size $L$, in order to study the scalability of the AVQITE approach. Since we are unable to determine \emph{a priori} which reference states produce the smallest $N_{CX}$, we have adopted the following procedure for obtaining the scaling data. We order the $L=4$ data as described in the previous paragraph and select the best $16$ reference states for the simulation of the larger system sizes. Note that we exclude unary encoding as it was clearly the worst in terms of gate count. 

\begin{figure*}
\begin{centering}
\includegraphics[width=\textwidth]{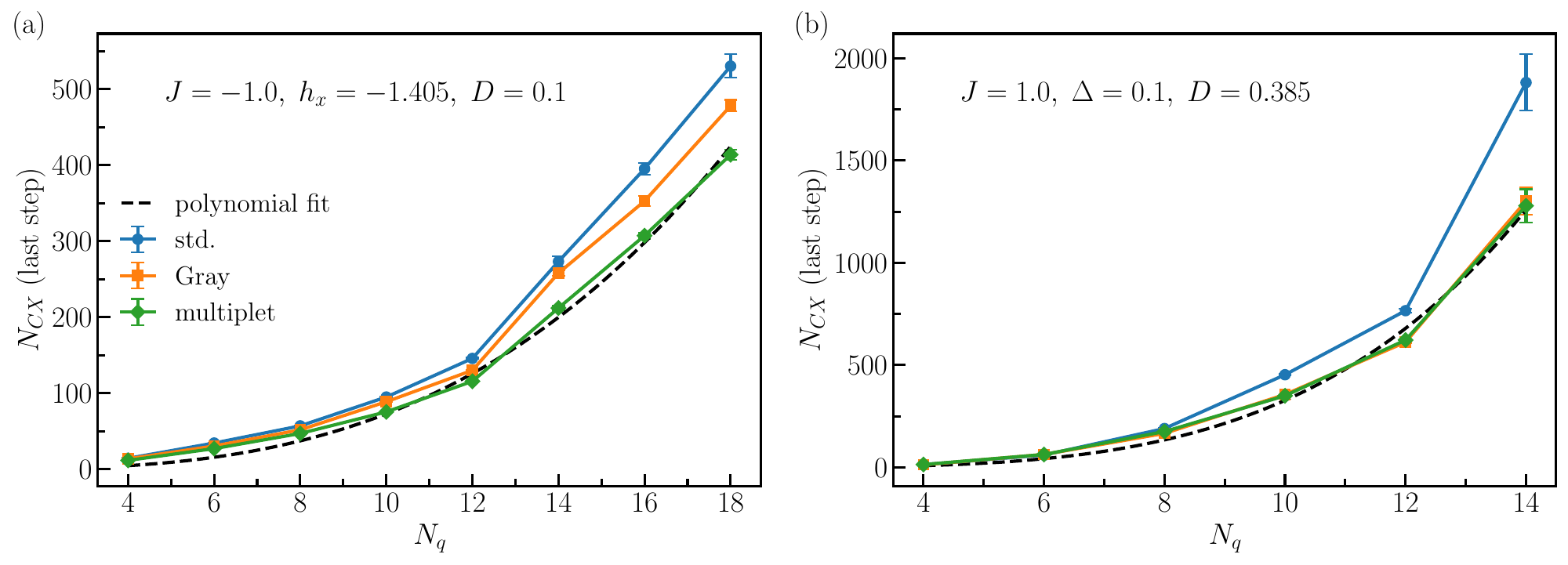}\caption{
	\label{fig:scaling} 
  Average number of CNOTs $N_{CX}$ required in the last step of the AVQITE routine as a function of number of qubits $N = 2L$ for (a) BC model and (b) XXZ models with parameters as given in the figure. The average is taken over a sample of 48 reference states (for details, see text). The dashed lines represent polynomial fits of the multiplet data using exponents $a L^3$ for panel (a) and $a' L^4$ for panel (b).}
\par\end{centering}
\end{figure*}

In Fig.~\ref{fig:scaling}, we show the scaling of $N_{CX}$ at the end of the simulation with the number of qubits $N = nL$ for the BC and XXZ models. Note that $n=2$ for the shown encodings. We find a cubic power-law scaling for the BC model, $N_{CX} \propto L^3$, whereas the XXZ models exhibits a scaling like $N_{CX} \propto L^4$. The dashed line in each panel denotes a power law fit to the multiplet encoded data with exponents $3$ and $4$, respectively. The multiplet and Gray code encodings exhibit a smaller power-law prefactor than the standard binary encoding for both models,  and are thus expected to give better performance in a real quantum device. When comparing multiplet and Gray encodings, the former shows a slight advantage for the BC model (Fig.~\ref{fig:scaling}\hyperlink{fig:scaling}{(a)}). For the XXZ model the two scaling curves are approximately identical.

\section{\label{sec:discussion}Conclusion}
In this work we have applied the AVQITE method to find ground states of spin-1 chains. As a benchmark, we have considered two well-known models, the Blume-Capel and XXZ models, which both contain single-ion anistropy terms characteristic of higher spin models. We used AVQITE to determine the ground state in the vicinity of a continuous quantum phase transition. We studied in detail the dependence of the convergence properties and the depth of the final AVQITE quantum circuit on the different hyperparameters of the algorithm: the choice of spin-to-qubit encoding, the operator pool, and the initial reference state. 

Using the total number of CNOT gates in the final quantum circuit as a figure of merit, we have found the unary encoding to be worse than the multiplet, Gray, and standard encodings, which yield comparable results. Averaging over all initial product states in the qubit $x$-and the $z$-basis, we have found the $z$-basis initial states with the maximal operator pool to perform best for both models at the selected parameter values. We have further identified that reference states that result in small $N_{CX}$ counts have a small initial energy variance, which can thus be used as an indicator for a good reference state. Further work is necessary to identify an optimal strategy for selecting the reference state. Finally, we have determined the scaling of $N_{CX}$ with system size and found $N_{CX} \propto L^3$ for the Blume-Capel model and $N_{CX} \propto L^4$ for the XXZ model. We note that a previous AVQITE study on spin-1/2 models reported $N_{CX} \propto L^2$~\cite{AVQITE}, which demonstrates that simulating spin-1 models requires more quantum resources. It would be interesting to study this scaling for spin models with higher $S > 1$. Other interesting future directions include applications of AVQITE to disordered and/or mixed-spin models. Finally, on the algorithmic side, the next important step is to test the impact of shot noise on the convergence properties of the AVQITE algorithm and to fully execute it on quantum hardware.

\section{\label{sec:thanks}Acknowledgments}

This work was supported by the U.S. Department of Energy (DOE), Office of Science, Basic Energy Sciences, 
Materials Science and Engineering Division, including the grant of computer time at the National Energy 
Research Scientific Computing Center (NERSC) in Berkeley, California. The research was performed at the 
Ames National Laboratory, which is operated for the U.S. DOE by Iowa State University under Contract No. DE-AC02-07CH11358. We thank Niladri Gomes and Wyatt Reller for useful discussions.

\bibliographystyle{apsrev4-2}
\bibliography{ref}

\begin{thebibliography}{77}%
\makeatletter
\providecommand \@ifxundefined [1]{%
 \@ifx{#1\undefined}
}%
\providecommand \@ifnum [1]{%
 \ifnum #1\expandafter \@firstoftwo
 \else \expandafter \@secondoftwo
 \fi
}%
\providecommand \@ifx [1]{%
 \ifx #1\expandafter \@firstoftwo
 \else \expandafter \@secondoftwo
 \fi
}%
\providecommand \natexlab [1]{#1}%
\providecommand \enquote  [1]{``#1''}%
\providecommand \bibnamefont  [1]{#1}%
\providecommand \bibfnamefont [1]{#1}%
\providecommand \citenamefont [1]{#1}%
\providecommand \href@noop [0]{\@secondoftwo}%
\providecommand \href [0]{\begingroup \@sanitize@url \@href}%
\providecommand \@href[1]{\@@startlink{#1}\@@href}%
\providecommand \@@href[1]{\endgroup#1\@@endlink}%
\providecommand \@sanitize@url [0]{\catcode `\\12\catcode `\$12\catcode
  `\&12\catcode `\#12\catcode `\^12\catcode `\_12\catcode `\%12\relax}%
\providecommand \@@startlink[1]{}%
\providecommand \@@endlink[0]{}%
\providecommand \url  [0]{\begingroup\@sanitize@url \@url }%
\providecommand \@url [1]{\endgroup\@href {#1}{\urlprefix }}%
\providecommand \urlprefix  [0]{URL }%
\providecommand \Eprint [0]{\href }%
\providecommand \doibase [0]{https://doi.org/}%
\providecommand \selectlanguage [0]{\@gobble}%
\providecommand \bibinfo  [0]{\@secondoftwo}%
\providecommand \bibfield  [0]{\@secondoftwo}%
\providecommand \translation [1]{[#1]}%
\providecommand \BibitemOpen [0]{}%
\providecommand \bibitemStop [0]{}%
\providecommand \bibitemNoStop [0]{.\EOS\space}%
\providecommand \EOS [0]{\spacefactor3000\relax}%
\providecommand \BibitemShut  [1]{\csname bibitem#1\endcsname}%
\let\auto@bib@innerbib\@empty
\bibitem [{\citenamefont {Aspuru-Guzik}\ \emph {et~al.}(2005)\citenamefont
  {Aspuru-Guzik}, \citenamefont {Dutoi}, \citenamefont {Love},\ and\
  \citenamefont {Head-Gordon}}]{asp_ipea}%
  \BibitemOpen
  \bibfield  {author} {\bibinfo {author} {\bibfnamefont {A.}~\bibnamefont
  {Aspuru-Guzik}}, \bibinfo {author} {\bibfnamefont {A.~D.}\ \bibnamefont
  {Dutoi}}, \bibinfo {author} {\bibfnamefont {P.~J.}\ \bibnamefont {Love}},\
  and\ \bibinfo {author} {\bibfnamefont {M.}~\bibnamefont {Head-Gordon}},\
  }\href {https://doi.org/10.1126/science.1113479} {\bibfield  {journal}
  {\bibinfo  {journal} {Science}\ }\textbf {\bibinfo {volume} {309}},\ \bibinfo
  {pages} {1704} (\bibinfo {year} {2005})}\BibitemShut {NoStop}%
\bibitem [{\citenamefont {Peruzzo}\ \emph {et~al.}(2014)\citenamefont
  {Peruzzo}, \citenamefont {McClean}, \citenamefont {Shadbolt}, \citenamefont
  {Yung}, \citenamefont {Zhou}, \citenamefont {Love}, \citenamefont
  {{Aspuru-Guzik}},\ and\ \citenamefont
  {O'Brien}}]{peruzzoVariationalEigenvalueSolver2014}%
  \BibitemOpen
  \bibfield  {author} {\bibinfo {author} {\bibfnamefont {A.}~\bibnamefont
  {Peruzzo}}, \bibinfo {author} {\bibfnamefont {J.}~\bibnamefont {McClean}},
  \bibinfo {author} {\bibfnamefont {P.}~\bibnamefont {Shadbolt}}, \bibinfo
  {author} {\bibfnamefont {M.-H.}\ \bibnamefont {Yung}}, \bibinfo {author}
  {\bibfnamefont {X.-Q.}\ \bibnamefont {Zhou}}, \bibinfo {author}
  {\bibfnamefont {P.~J.}\ \bibnamefont {Love}}, \bibinfo {author}
  {\bibfnamefont {A.}~\bibnamefont {{Aspuru-Guzik}}},\ and\ \bibinfo {author}
  {\bibfnamefont {J.~L.}\ \bibnamefont {O'Brien}},\ }\href
  {https://doi.org/10.1038/ncomms5213} {\bibfield  {journal} {\bibinfo
  {journal} {Nat. Commun.}\ }\textbf {\bibinfo {volume} {5}},\ \bibinfo {pages}
  {1} (\bibinfo {year} {2014})}\BibitemShut {NoStop}%
\bibitem [{\citenamefont {Kandala}\ \emph {et~al.}(2017)\citenamefont
  {Kandala}, \citenamefont {Mezzacapo}, \citenamefont {Temme}, \citenamefont
  {Takita}, \citenamefont {Brink}, \citenamefont {Chow},\ and\ \citenamefont
  {Gambetta}}]{hardware_efficient_vqe}%
  \BibitemOpen
  \bibfield  {author} {\bibinfo {author} {\bibfnamefont {A.}~\bibnamefont
  {Kandala}}, \bibinfo {author} {\bibfnamefont {A.}~\bibnamefont {Mezzacapo}},
  \bibinfo {author} {\bibfnamefont {K.}~\bibnamefont {Temme}}, \bibinfo
  {author} {\bibfnamefont {M.}~\bibnamefont {Takita}}, \bibinfo {author}
  {\bibfnamefont {M.}~\bibnamefont {Brink}}, \bibinfo {author} {\bibfnamefont
  {J.~M.}\ \bibnamefont {Chow}},\ and\ \bibinfo {author} {\bibfnamefont
  {J.~M.}\ \bibnamefont {Gambetta}},\ }\href
  {https://doi.org/10.1038/nature23879} {\bibfield  {journal} {\bibinfo
  {journal} {Nature}\ }\textbf {\bibinfo {volume} {549}},\ \bibinfo {pages}
  {242} (\bibinfo {year} {2017})}\BibitemShut {NoStop}%
\bibitem [{\citenamefont {O’Malley}\ \emph {et~al.}(2016)\citenamefont
  {O’Malley}, \citenamefont {Babbush}, \citenamefont {Kivlichan},
  \citenamefont {Romero}, \citenamefont {McClean}, \citenamefont {Barends},
  \citenamefont {Kelly}, \citenamefont {Roushan}, \citenamefont {Tranter},
  \citenamefont {Ding} \emph {et~al.}}]{vqe_pea_h2}%
  \BibitemOpen
  \bibfield  {author} {\bibinfo {author} {\bibfnamefont {P.~J.}\ \bibnamefont
  {O’Malley}}, \bibinfo {author} {\bibfnamefont {R.}~\bibnamefont {Babbush}},
  \bibinfo {author} {\bibfnamefont {I.~D.}\ \bibnamefont {Kivlichan}}, \bibinfo
  {author} {\bibfnamefont {J.}~\bibnamefont {Romero}}, \bibinfo {author}
  {\bibfnamefont {J.~R.}\ \bibnamefont {McClean}}, \bibinfo {author}
  {\bibfnamefont {R.}~\bibnamefont {Barends}}, \bibinfo {author} {\bibfnamefont
  {J.}~\bibnamefont {Kelly}}, \bibinfo {author} {\bibfnamefont
  {P.}~\bibnamefont {Roushan}}, \bibinfo {author} {\bibfnamefont
  {A.}~\bibnamefont {Tranter}}, \bibinfo {author} {\bibfnamefont
  {N.}~\bibnamefont {Ding}}, \emph {et~al.},\ }\href
  {https://doi.org/10.1103/PhysRevX.6.031007} {\bibfield  {journal} {\bibinfo
  {journal} {Phys. Rev. X}\ }\textbf {\bibinfo {volume} {6}},\ \bibinfo {pages}
  {031007} (\bibinfo {year} {2016})}\BibitemShut {NoStop}%
\bibitem [{\citenamefont {Preskill}(2018)}]{nisq}%
  \BibitemOpen
  \bibfield  {author} {\bibinfo {author} {\bibfnamefont {J.}~\bibnamefont
  {Preskill}},\ }\href {https://doi.org/10.22331/q-2018-08-06-79} {\bibfield
  {journal} {\bibinfo  {journal} {Quantum}\ }\textbf {\bibinfo {volume} {2}},\
  \bibinfo {pages} {79} (\bibinfo {year} {2018})}\BibitemShut {NoStop}%
\bibitem [{\citenamefont {McArdle}\ \emph {et~al.}(2020)\citenamefont
  {McArdle}, \citenamefont {Endo}, \citenamefont {Aspuru-Guzik}, \citenamefont
  {Benjamin},\ and\ \citenamefont {Yuan}}]{rmp_qcc}%
  \BibitemOpen
  \bibfield  {author} {\bibinfo {author} {\bibfnamefont {S.}~\bibnamefont
  {McArdle}}, \bibinfo {author} {\bibfnamefont {S.}~\bibnamefont {Endo}},
  \bibinfo {author} {\bibfnamefont {A.}~\bibnamefont {Aspuru-Guzik}}, \bibinfo
  {author} {\bibfnamefont {S.~C.}\ \bibnamefont {Benjamin}},\ and\ \bibinfo
  {author} {\bibfnamefont {X.}~\bibnamefont {Yuan}},\ }\href
  {https://doi.org/10.1103/RevModPhys.92.015003} {\bibfield  {journal}
  {\bibinfo  {journal} {Rev. Mod. Phys.}\ }\textbf {\bibinfo {volume} {92}},\
  \bibinfo {pages} {015003} (\bibinfo {year} {2020})}\BibitemShut {NoStop}%
\bibitem [{\citenamefont {Bauer}\ \emph {et~al.}(2020)\citenamefont {Bauer},
  \citenamefont {Bravyi}, \citenamefont {Motta},\ and\ \citenamefont
  {Chan}}]{bauer2020quantum}%
  \BibitemOpen
  \bibfield  {author} {\bibinfo {author} {\bibfnamefont {B.}~\bibnamefont
  {Bauer}}, \bibinfo {author} {\bibfnamefont {S.}~\bibnamefont {Bravyi}},
  \bibinfo {author} {\bibfnamefont {M.}~\bibnamefont {Motta}},\ and\ \bibinfo
  {author} {\bibfnamefont {G.~K.-L.}\ \bibnamefont {Chan}},\ }\href
  {https://doi.org/10.1021/acs.chemrev.9b00829} {\bibfield  {journal} {\bibinfo
   {journal} {Chem. Rev.}\ }\textbf {\bibinfo {volume} {120}},\ \bibinfo
  {pages} {12685} (\bibinfo {year} {2020})}\BibitemShut {NoStop}%
\bibitem [{\citenamefont {Cerezo}\ \emph {et~al.}(2021)\citenamefont {Cerezo},
  \citenamefont {Arrasmith}, \citenamefont {Babbush}, \citenamefont {Benjamin},
  \citenamefont {Endo}, \citenamefont {Fujii}, \citenamefont {McClean},
  \citenamefont {Mitarai}, \citenamefont {Yuan}, \citenamefont {Cincio} \emph
  {et~al.}}]{cerezo2021variational}%
  \BibitemOpen
  \bibfield  {author} {\bibinfo {author} {\bibfnamefont {M.}~\bibnamefont
  {Cerezo}}, \bibinfo {author} {\bibfnamefont {A.}~\bibnamefont {Arrasmith}},
  \bibinfo {author} {\bibfnamefont {R.}~\bibnamefont {Babbush}}, \bibinfo
  {author} {\bibfnamefont {S.~C.}\ \bibnamefont {Benjamin}}, \bibinfo {author}
  {\bibfnamefont {S.}~\bibnamefont {Endo}}, \bibinfo {author} {\bibfnamefont
  {K.}~\bibnamefont {Fujii}}, \bibinfo {author} {\bibfnamefont {J.~R.}\
  \bibnamefont {McClean}}, \bibinfo {author} {\bibfnamefont {K.}~\bibnamefont
  {Mitarai}}, \bibinfo {author} {\bibfnamefont {X.}~\bibnamefont {Yuan}},
  \bibinfo {author} {\bibfnamefont {L.}~\bibnamefont {Cincio}}, \emph
  {et~al.},\ }\href {https://doi.org/10.1038/s42254-021-00348-9} {\bibfield
  {journal} {\bibinfo  {journal} {Nat. Rev. Phys.}\ }\textbf {\bibinfo {volume}
  {3}},\ \bibinfo {pages} {625} (\bibinfo {year} {2021})}\BibitemShut {NoStop}%
\bibitem [{\citenamefont {Motta}\ \emph {et~al.}(2020)\citenamefont {Motta},
  \citenamefont {Sun}, \citenamefont {Tan}, \citenamefont {O’Rourke},
  \citenamefont {Ye}, \citenamefont {Minnich}, \citenamefont {Brand{\~a}o},\
  and\ \citenamefont {Chan}}]{qite_chan20}%
  \BibitemOpen
  \bibfield  {author} {\bibinfo {author} {\bibfnamefont {M.}~\bibnamefont
  {Motta}}, \bibinfo {author} {\bibfnamefont {C.}~\bibnamefont {Sun}}, \bibinfo
  {author} {\bibfnamefont {A.~T.}\ \bibnamefont {Tan}}, \bibinfo {author}
  {\bibfnamefont {M.~J.}\ \bibnamefont {O’Rourke}}, \bibinfo {author}
  {\bibfnamefont {E.}~\bibnamefont {Ye}}, \bibinfo {author} {\bibfnamefont
  {A.~J.}\ \bibnamefont {Minnich}}, \bibinfo {author} {\bibfnamefont {F.~G.}\
  \bibnamefont {Brand{\~a}o}},\ and\ \bibinfo {author} {\bibfnamefont
  {G.~K.-L.}\ \bibnamefont {Chan}},\ }\href
  {https://doi.org/10.1038/s41567-019-0704-4} {\bibfield  {journal} {\bibinfo
  {journal} {Nat. Phys.}\ }\textbf {\bibinfo {volume} {16}},\ \bibinfo {pages}
  {205} (\bibinfo {year} {2020})}\BibitemShut {NoStop}%
\bibitem [{\citenamefont {Berthusen}\ \emph {et~al.}(2022)\citenamefont
  {Berthusen}, \citenamefont {Trevisan}, \citenamefont {Iadecola},\ and\
  \citenamefont {Orth}}]{Berthusen:2022}%
  \BibitemOpen
  \bibfield  {author} {\bibinfo {author} {\bibfnamefont {N.~F.}\ \bibnamefont
  {Berthusen}}, \bibinfo {author} {\bibfnamefont {T.~V.}\ \bibnamefont
  {Trevisan}}, \bibinfo {author} {\bibfnamefont {T.}~\bibnamefont {Iadecola}},\
  and\ \bibinfo {author} {\bibfnamefont {P.~P.}\ \bibnamefont {Orth}},\ }\href
  {https://doi.org/10.1103/PhysRevResearch.4.023097} {\bibfield  {journal}
  {\bibinfo  {journal} {Phys. Rev. Res.}\ }\textbf {\bibinfo {volume} {4}},\
  \bibinfo {pages} {023097} (\bibinfo {year} {2022})}\BibitemShut {NoStop}%
\bibitem [{\citenamefont {Chen}\ \emph {et~al.}(2022)\citenamefont {Chen},
  \citenamefont {Burdick}, \citenamefont {Yao}, \citenamefont {Orth},\ and\
  \citenamefont {Iadecola}}]{Chen2022}%
  \BibitemOpen
  \bibfield  {author} {\bibinfo {author} {\bibfnamefont {I.-C.}\ \bibnamefont
  {Chen}}, \bibinfo {author} {\bibfnamefont {B.}~\bibnamefont {Burdick}},
  \bibinfo {author} {\bibfnamefont {Y.}~\bibnamefont {Yao}}, \bibinfo {author}
  {\bibfnamefont {P.~P.}\ \bibnamefont {Orth}},\ and\ \bibinfo {author}
  {\bibfnamefont {T.}~\bibnamefont {Iadecola}},\ }\href
  {https://doi.org/10.1103/PhysRevResearch.4.043027} {\bibfield  {journal}
  {\bibinfo  {journal} {Phys. Rev. Res.}\ }\textbf {\bibinfo {volume} {4}},\
  \bibinfo {pages} {043027} (\bibinfo {year} {2022})}\BibitemShut {NoStop}%
\bibitem [{\citenamefont {Tilly}\ \emph {et~al.}(2022)\citenamefont {Tilly},
  \citenamefont {Chen}, \citenamefont {Cao}, \citenamefont {Picozzi},
  \citenamefont {Setia}, \citenamefont {Li}, \citenamefont {Grant},
  \citenamefont {Wossnig}, \citenamefont {Rungger}, \citenamefont {Booth} \emph
  {et~al.}}]{tilly2022variational}%
  \BibitemOpen
  \bibfield  {author} {\bibinfo {author} {\bibfnamefont {J.}~\bibnamefont
  {Tilly}}, \bibinfo {author} {\bibfnamefont {H.}~\bibnamefont {Chen}},
  \bibinfo {author} {\bibfnamefont {S.}~\bibnamefont {Cao}}, \bibinfo {author}
  {\bibfnamefont {D.}~\bibnamefont {Picozzi}}, \bibinfo {author} {\bibfnamefont
  {K.}~\bibnamefont {Setia}}, \bibinfo {author} {\bibfnamefont
  {Y.}~\bibnamefont {Li}}, \bibinfo {author} {\bibfnamefont {E.}~\bibnamefont
  {Grant}}, \bibinfo {author} {\bibfnamefont {L.}~\bibnamefont {Wossnig}},
  \bibinfo {author} {\bibfnamefont {I.}~\bibnamefont {Rungger}}, \bibinfo
  {author} {\bibfnamefont {G.~H.}\ \bibnamefont {Booth}}, \emph {et~al.},\
  }\href {https://doi.org/10.1016/j.physrep.2022.08.003} {\bibfield  {journal}
  {\bibinfo  {journal} {Phys. Rep.}\ }\textbf {\bibinfo {volume} {986}},\
  \bibinfo {pages} {1} (\bibinfo {year} {2022})}\BibitemShut {NoStop}%
\bibitem [{\citenamefont {{Huggins}}\ \emph {et~al.}(2022)\citenamefont
  {{Huggins}}, \citenamefont {{O'Gorman}}, \citenamefont {{Rubin}},
  \citenamefont {{Reichman}}, \citenamefont {{Babbush}},\ and\ \citenamefont
  {{Lee}}}]{Huggins2022Unbiasingfq}%
  \BibitemOpen
  \bibfield  {author} {\bibinfo {author} {\bibfnamefont {W.~J.}\ \bibnamefont
  {{Huggins}}}, \bibinfo {author} {\bibfnamefont {B.~A.}\ \bibnamefont
  {{O'Gorman}}}, \bibinfo {author} {\bibfnamefont {N.~C.}\ \bibnamefont
  {{Rubin}}}, \bibinfo {author} {\bibfnamefont {D.~R.}\ \bibnamefont
  {{Reichman}}}, \bibinfo {author} {\bibfnamefont {R.}~\bibnamefont
  {{Babbush}}},\ and\ \bibinfo {author} {\bibfnamefont {J.}~\bibnamefont
  {{Lee}}},\ }\href {https://doi.org/10.1038/s41586-021-04351-z} {\bibfield
  {journal} {\bibinfo  {journal} {Nature}\ }\textbf {\bibinfo {volume} {603}},\
  \bibinfo {pages} {416} (\bibinfo {year} {2022})}\BibitemShut {NoStop}%
\bibitem [{\citenamefont {Kim}\ \emph {et~al.}(2023{\natexlab{a}})\citenamefont
  {Kim}, \citenamefont {Wood}, \citenamefont {Yoder}, \citenamefont {Merkel},
  \citenamefont {Gambetta}, \citenamefont {Temme},\ and\ \citenamefont
  {Kandala}}]{kim2023scalable}%
  \BibitemOpen
  \bibfield  {author} {\bibinfo {author} {\bibfnamefont {Y.}~\bibnamefont
  {Kim}}, \bibinfo {author} {\bibfnamefont {C.~J.}\ \bibnamefont {Wood}},
  \bibinfo {author} {\bibfnamefont {T.~J.}\ \bibnamefont {Yoder}}, \bibinfo
  {author} {\bibfnamefont {S.~T.}\ \bibnamefont {Merkel}}, \bibinfo {author}
  {\bibfnamefont {J.~M.}\ \bibnamefont {Gambetta}}, \bibinfo {author}
  {\bibfnamefont {K.}~\bibnamefont {Temme}},\ and\ \bibinfo {author}
  {\bibfnamefont {A.}~\bibnamefont {Kandala}},\ }\href
  {https://doi.org/10.1038/s41567-022-01914-3} {\bibfield  {journal} {\bibinfo
  {journal} {Nat. Phys.}\ ,\ \bibinfo {pages} {1}} (\bibinfo {year}
  {2023}{\natexlab{a}})}\BibitemShut {NoStop}%
\bibitem [{\citenamefont {{van den Berg}}\ \emph {et~al.}(2023)\citenamefont
  {{van den Berg}}, \citenamefont {Minev}, \citenamefont {Kandala},\ and\
  \citenamefont {Temme}}]{vandenbergProbabilisticErrorCancellation2023}%
  \BibitemOpen
  \bibfield  {author} {\bibinfo {author} {\bibfnamefont {E.}~\bibnamefont {{van
  den Berg}}}, \bibinfo {author} {\bibfnamefont {Z.~K.}\ \bibnamefont {Minev}},
  \bibinfo {author} {\bibfnamefont {A.}~\bibnamefont {Kandala}},\ and\ \bibinfo
  {author} {\bibfnamefont {K.}~\bibnamefont {Temme}},\ }\href
  {https://doi.org/10.1038/s41567-023-02042-2} {\bibfield  {journal} {\bibinfo
  {journal} {Nat. Phys.}\ ,\ \bibinfo {pages} {1}} (\bibinfo {year}
  {2023})}\BibitemShut {NoStop}%
\bibitem [{\citenamefont {Kim}\ \emph {et~al.}(2023{\natexlab{b}})\citenamefont
  {Kim}, \citenamefont {Eddins}, \citenamefont {Anand}, \citenamefont {Wei},
  \citenamefont {{van den Berg}}, \citenamefont {Rosenblatt}, \citenamefont
  {Nayfeh}, \citenamefont {Wu}, \citenamefont {Zaletel}, \citenamefont
  {Temme},\ and\ \citenamefont {Kandala}}]{kimEvidenceUtilityQuantum2023}%
  \BibitemOpen
  \bibfield  {author} {\bibinfo {author} {\bibfnamefont {Y.}~\bibnamefont
  {Kim}}, \bibinfo {author} {\bibfnamefont {A.}~\bibnamefont {Eddins}},
  \bibinfo {author} {\bibfnamefont {S.}~\bibnamefont {Anand}}, \bibinfo
  {author} {\bibfnamefont {K.~X.}\ \bibnamefont {Wei}}, \bibinfo {author}
  {\bibfnamefont {E.}~\bibnamefont {{van den Berg}}}, \bibinfo {author}
  {\bibfnamefont {S.}~\bibnamefont {Rosenblatt}}, \bibinfo {author}
  {\bibfnamefont {H.}~\bibnamefont {Nayfeh}}, \bibinfo {author} {\bibfnamefont
  {Y.}~\bibnamefont {Wu}}, \bibinfo {author} {\bibfnamefont {M.}~\bibnamefont
  {Zaletel}}, \bibinfo {author} {\bibfnamefont {K.}~\bibnamefont {Temme}},\
  and\ \bibinfo {author} {\bibfnamefont {A.}~\bibnamefont {Kandala}},\ }\href
  {https://doi.org/10.1038/s41586-023-06096-3} {\bibfield  {journal} {\bibinfo
  {journal} {Nature}\ }\textbf {\bibinfo {volume} {618}},\ \bibinfo {pages}
  {500} (\bibinfo {year} {2023}{\natexlab{b}})}\BibitemShut {NoStop}%
\bibitem [{\citenamefont {Setia}\ \emph {et~al.}(2019)\citenamefont {Setia},
  \citenamefont {Bravyi}, \citenamefont {Mezzacapo},\ and\ \citenamefont
  {Whitfield}}]{setia2019}%
  \BibitemOpen
  \bibfield  {author} {\bibinfo {author} {\bibfnamefont {K.}~\bibnamefont
  {Setia}}, \bibinfo {author} {\bibfnamefont {S.}~\bibnamefont {Bravyi}},
  \bibinfo {author} {\bibfnamefont {A.}~\bibnamefont {Mezzacapo}},\ and\
  \bibinfo {author} {\bibfnamefont {J.~D.}\ \bibnamefont {Whitfield}},\ }\href
  {https://doi.org/10.1103/PhysRevResearch.1.033033} {\bibfield  {journal}
  {\bibinfo  {journal} {Phys. Rev. Res.}\ }\textbf {\bibinfo {volume} {1}},\
  \bibinfo {pages} {033033} (\bibinfo {year} {2019})}\BibitemShut {NoStop}%
\bibitem [{\citenamefont {Getelina}\ \emph {et~al.}(2023)\citenamefont
  {Getelina}, \citenamefont {Gomes}, \citenamefont {Iadecola}, \citenamefont
  {Orth},\ and\ \citenamefont {Yao}}]{getelina2023adaptive}%
  \BibitemOpen
  \bibfield  {author} {\bibinfo {author} {\bibfnamefont {J.~C.}\ \bibnamefont
  {Getelina}}, \bibinfo {author} {\bibfnamefont {N.}~\bibnamefont {Gomes}},
  \bibinfo {author} {\bibfnamefont {T.}~\bibnamefont {Iadecola}}, \bibinfo
  {author} {\bibfnamefont {P.~P.}\ \bibnamefont {Orth}},\ and\ \bibinfo
  {author} {\bibfnamefont {Y.-X.}\ \bibnamefont {Yao}},\ }\href
  {https://doi.org/10.21468/SciPostPhys.15.3.102} {\bibfield  {journal}
  {\bibinfo  {journal} {SciPost Phys.}\ }\textbf {\bibinfo {volume} {15}},\
  \bibinfo {pages} {102} (\bibinfo {year} {2023})}\BibitemShut {NoStop}%
\bibitem [{\citenamefont {Miessen}\ \emph {et~al.}(2023)\citenamefont
  {Miessen}, \citenamefont {Ollitrault}, \citenamefont {Tacchino},\ and\
  \citenamefont {Tavernelli}}]{miessen2023quantum}%
  \BibitemOpen
  \bibfield  {author} {\bibinfo {author} {\bibfnamefont {A.}~\bibnamefont
  {Miessen}}, \bibinfo {author} {\bibfnamefont {P.~J.}\ \bibnamefont
  {Ollitrault}}, \bibinfo {author} {\bibfnamefont {F.}~\bibnamefont
  {Tacchino}},\ and\ \bibinfo {author} {\bibfnamefont {I.}~\bibnamefont
  {Tavernelli}},\ }\href {https://doi.org/10.1038/s43588-022-00374-2}
  {\bibfield  {journal} {\bibinfo  {journal} {Nat. Comput. Sci.}\ }\textbf
  {\bibinfo {volume} {3}},\ \bibinfo {pages} {25} (\bibinfo {year}
  {2023})}\BibitemShut {NoStop}%
\bibitem [{\citenamefont {Shtanko}\ \emph {et~al.}()\citenamefont {Shtanko},
  \citenamefont {Wang}, \citenamefont {Zhang}, \citenamefont {Harle},
  \citenamefont {Seif}, \citenamefont {Movassagh},\ and\ \citenamefont
  {Minev}}]{shtanko2023uncovering}%
  \BibitemOpen
  \bibfield  {author} {\bibinfo {author} {\bibfnamefont {O.}~\bibnamefont
  {Shtanko}}, \bibinfo {author} {\bibfnamefont {D.~S.}\ \bibnamefont {Wang}},
  \bibinfo {author} {\bibfnamefont {H.}~\bibnamefont {Zhang}}, \bibinfo
  {author} {\bibfnamefont {N.}~\bibnamefont {Harle}}, \bibinfo {author}
  {\bibfnamefont {A.}~\bibnamefont {Seif}}, \bibinfo {author} {\bibfnamefont
  {R.}~\bibnamefont {Movassagh}},\ and\ \bibinfo {author} {\bibfnamefont
  {Z.}~\bibnamefont {Minev}},\ }\href@noop {} {}\Eprint
  {https://arxiv.org/abs/2307.07552} {arXiv:2307.07552} \BibitemShut {NoStop}%
\bibitem [{\citenamefont {{Ogunkoya}}\ \emph {et~al.}()\citenamefont
  {{Ogunkoya}}, \citenamefont {{Kim}}, \citenamefont {{Peng}}, \citenamefont
  {{Bar{\i}{\c{s}} {\"O}zg{\"u}ler}},\ and\ \citenamefont
  {{Alexeev}}}]{ogunkoya2023qutrit}%
  \BibitemOpen
  \bibfield  {author} {\bibinfo {author} {\bibfnamefont {O.}~\bibnamefont
  {{Ogunkoya}}}, \bibinfo {author} {\bibfnamefont {J.}~\bibnamefont {{Kim}}},
  \bibinfo {author} {\bibfnamefont {B.}~\bibnamefont {{Peng}}}, \bibinfo
  {author} {\bibfnamefont {A.}~\bibnamefont {{Bar{\i}{\c{s}}
  {\"O}zg{\"u}ler}}},\ and\ \bibinfo {author} {\bibfnamefont {Y.}~\bibnamefont
  {{Alexeev}}},\ }\href@noop {} {}\Eprint {https://arxiv.org/abs/2309.00740}
  {arXiv:2309.00740} \BibitemShut {NoStop}%
\bibitem [{\citenamefont {Wang}\ \emph {et~al.}(2023)\citenamefont {Wang},
  \citenamefont {Snizhko}, \citenamefont {Romito}, \citenamefont {Gefen},\ and\
  \citenamefont {Murch}}]{wang2023}%
  \BibitemOpen
  \bibfield  {author} {\bibinfo {author} {\bibfnamefont {Y.}~\bibnamefont
  {Wang}}, \bibinfo {author} {\bibfnamefont {K.}~\bibnamefont {Snizhko}},
  \bibinfo {author} {\bibfnamefont {A.}~\bibnamefont {Romito}}, \bibinfo
  {author} {\bibfnamefont {Y.}~\bibnamefont {Gefen}},\ and\ \bibinfo {author}
  {\bibfnamefont {K.}~\bibnamefont {Murch}},\ }\href
  {https://doi.org/10.1103/PhysRevA.108.013712} {\bibfield  {journal} {\bibinfo
   {journal} {Phys. Rev. A}\ }\textbf {\bibinfo {volume} {108}},\ \bibinfo
  {pages} {013712} (\bibinfo {year} {2023})}\BibitemShut {NoStop}%
\bibitem [{\citenamefont {Haldane}(1983{\natexlab{a}})}]{haldane1983a}%
  \BibitemOpen
  \bibfield  {author} {\bibinfo {author} {\bibfnamefont {F.~D.~M.}\
  \bibnamefont {Haldane}},\ }\href
  {https://doi.org/10.1103/PhysRevLett.50.1153} {\bibfield  {journal} {\bibinfo
   {journal} {Phys. Rev. Lett.}\ }\textbf {\bibinfo {volume} {50}},\ \bibinfo
  {pages} {1153} (\bibinfo {year} {1983}{\natexlab{a}})}\BibitemShut {NoStop}%
\bibitem [{\citenamefont {Haldane}(1983{\natexlab{b}})}]{haldane1983b}%
  \BibitemOpen
  \bibfield  {author} {\bibinfo {author} {\bibfnamefont {F.}~\bibnamefont
  {Haldane}},\ }\href {https://doi.org/10.1016/0375-9601(83)90631-X} {\bibfield
   {journal} {\bibinfo  {journal} {Phys. Lett. A}\ }\textbf {\bibinfo {volume}
  {93}},\ \bibinfo {pages} {464} (\bibinfo {year}
  {1983}{\natexlab{b}})}\BibitemShut {NoStop}%
\bibitem [{\citenamefont {Lieb}\ \emph {et~al.}(1961)\citenamefont {Lieb},
  \citenamefont {Schultz},\ and\ \citenamefont {Mattis}}]{LIEB1961407}%
  \BibitemOpen
  \bibfield  {author} {\bibinfo {author} {\bibfnamefont {E.}~\bibnamefont
  {Lieb}}, \bibinfo {author} {\bibfnamefont {T.}~\bibnamefont {Schultz}},\ and\
  \bibinfo {author} {\bibfnamefont {D.}~\bibnamefont {Mattis}},\ }\href
  {https://doi.org/10.1016/0003-4916(61)90115-4} {\bibfield  {journal}
  {\bibinfo  {journal} {Annals of Physics}\ }\textbf {\bibinfo {volume} {16}},\
  \bibinfo {pages} {407} (\bibinfo {year} {1961})}\BibitemShut {NoStop}%
\bibitem [{\citenamefont {Affleck}\ \emph {et~al.}(1988)\citenamefont
  {Affleck}, \citenamefont {Kennedy}, \citenamefont {Lieb},\ and\ \citenamefont
  {Tasaki}}]{affleck1988}%
  \BibitemOpen
  \bibfield  {author} {\bibinfo {author} {\bibfnamefont {I.}~\bibnamefont
  {Affleck}}, \bibinfo {author} {\bibfnamefont {T.}~\bibnamefont {Kennedy}},
  \bibinfo {author} {\bibfnamefont {E.~H.}\ \bibnamefont {Lieb}},\ and\
  \bibinfo {author} {\bibfnamefont {H.}~\bibnamefont {Tasaki}},\ }\href
  {https://doi.org/10.1007/BF01218021} {\bibfield  {journal} {\bibinfo
  {journal} {Communications in Mathematical Physics}\ }\textbf {\bibinfo
  {volume} {115}},\ \bibinfo {pages} {477} (\bibinfo {year}
  {1988})}\BibitemShut {NoStop}%
\bibitem [{\citenamefont {Pollmann}\ \emph {et~al.}(2010)\citenamefont
  {Pollmann}, \citenamefont {Turner}, \citenamefont {Berg},\ and\ \citenamefont
  {Oshikawa}}]{pollmann2010}%
  \BibitemOpen
  \bibfield  {author} {\bibinfo {author} {\bibfnamefont {F.}~\bibnamefont
  {Pollmann}}, \bibinfo {author} {\bibfnamefont {A.~M.}\ \bibnamefont
  {Turner}}, \bibinfo {author} {\bibfnamefont {E.}~\bibnamefont {Berg}},\ and\
  \bibinfo {author} {\bibfnamefont {M.}~\bibnamefont {Oshikawa}},\ }\href
  {https://doi.org/10.1103/PhysRevB.81.064439} {\bibfield  {journal} {\bibinfo
  {journal} {Phys. Rev. B}\ }\textbf {\bibinfo {volume} {81}},\ \bibinfo
  {pages} {064439} (\bibinfo {year} {2010})}\BibitemShut {NoStop}%
\bibitem [{\citenamefont {Pollmann}\ \emph {et~al.}(2012)\citenamefont
  {Pollmann}, \citenamefont {Berg}, \citenamefont {Turner},\ and\ \citenamefont
  {Oshikawa}}]{polmann2012}%
  \BibitemOpen
  \bibfield  {author} {\bibinfo {author} {\bibfnamefont {F.}~\bibnamefont
  {Pollmann}}, \bibinfo {author} {\bibfnamefont {E.}~\bibnamefont {Berg}},
  \bibinfo {author} {\bibfnamefont {A.~M.}\ \bibnamefont {Turner}},\ and\
  \bibinfo {author} {\bibfnamefont {M.}~\bibnamefont {Oshikawa}},\ }\href
  {https://doi.org/10.1103/PhysRevB.85.075125} {\bibfield  {journal} {\bibinfo
  {journal} {Phys. Rev. B}\ }\textbf {\bibinfo {volume} {85}},\ \bibinfo
  {pages} {075125} (\bibinfo {year} {2012})}\BibitemShut {NoStop}%
\bibitem [{\citenamefont {Papanicolaou}(1988)}]{papanicolaou1988}%
  \BibitemOpen
  \bibfield  {author} {\bibinfo {author} {\bibfnamefont {N.}~\bibnamefont
  {Papanicolaou}},\ }\href
  {https://doi.org/https://doi.org/10.1016/0550-3213(88)90073-9} {\bibfield
  {journal} {\bibinfo  {journal} {Nuclear Physics B}\ }\textbf {\bibinfo
  {volume} {305}},\ \bibinfo {pages} {367} (\bibinfo {year}
  {1988})}\BibitemShut {NoStop}%
\bibitem [{\citenamefont {Harada}\ and\ \citenamefont
  {Kawashima}(2002)}]{harada2002}%
  \BibitemOpen
  \bibfield  {author} {\bibinfo {author} {\bibfnamefont {K.}~\bibnamefont
  {Harada}}\ and\ \bibinfo {author} {\bibfnamefont {N.}~\bibnamefont
  {Kawashima}},\ }\href {https://doi.org/10.1103/PhysRevB.65.052403} {\bibfield
   {journal} {\bibinfo  {journal} {Phys. Rev. B}\ }\textbf {\bibinfo {volume}
  {65}},\ \bibinfo {pages} {052403} (\bibinfo {year} {2002})}\BibitemShut
  {NoStop}%
\bibitem [{\citenamefont {Nakatsuji}\ \emph {et~al.}(2005)\citenamefont
  {Nakatsuji}, \citenamefont {Nambu}, \citenamefont {Tonomura}, \citenamefont
  {Sakai}, \citenamefont {Jonas}, \citenamefont {Broholm}, \citenamefont
  {Tsunetsugu}, \citenamefont {Qiu},\ and\ \citenamefont
  {Maeno}}]{broholm2005}%
  \BibitemOpen
  \bibfield  {author} {\bibinfo {author} {\bibfnamefont {S.}~\bibnamefont
  {Nakatsuji}}, \bibinfo {author} {\bibfnamefont {Y.}~\bibnamefont {Nambu}},
  \bibinfo {author} {\bibfnamefont {H.}~\bibnamefont {Tonomura}}, \bibinfo
  {author} {\bibfnamefont {O.}~\bibnamefont {Sakai}}, \bibinfo {author}
  {\bibfnamefont {S.}~\bibnamefont {Jonas}}, \bibinfo {author} {\bibfnamefont
  {C.}~\bibnamefont {Broholm}}, \bibinfo {author} {\bibfnamefont
  {H.}~\bibnamefont {Tsunetsugu}}, \bibinfo {author} {\bibfnamefont
  {Y.}~\bibnamefont {Qiu}},\ and\ \bibinfo {author} {\bibfnamefont
  {Y.}~\bibnamefont {Maeno}},\ }\href {https://doi.org/10.1126/science.1114727}
  {\bibfield  {journal} {\bibinfo  {journal} {Science}\ }\textbf {\bibinfo
  {volume} {309}},\ \bibinfo {pages} {1697} (\bibinfo {year}
  {2005})}\BibitemShut {NoStop}%
\bibitem [{\citenamefont {Tsunetsugu}\ and\ \citenamefont
  {Arikawa}(2006)}]{arikawa2006}%
  \BibitemOpen
  \bibfield  {author} {\bibinfo {author} {\bibfnamefont {H.}~\bibnamefont
  {Tsunetsugu}}\ and\ \bibinfo {author} {\bibfnamefont {M.}~\bibnamefont
  {Arikawa}},\ }\href {https://doi.org/10.1143/JPSJ.75.083701} {\bibfield
  {journal} {\bibinfo  {journal} {Journal of the Physical Society of Japan}\
  }\textbf {\bibinfo {volume} {75}},\ \bibinfo {pages} {083701} (\bibinfo
  {year} {2006})}\BibitemShut {NoStop}%
\bibitem [{\citenamefont {L\"auchli}\ \emph {et~al.}(2006)\citenamefont
  {L\"auchli}, \citenamefont {Mila},\ and\ \citenamefont
  {Penc}}]{laeuchli2006}%
  \BibitemOpen
  \bibfield  {author} {\bibinfo {author} {\bibfnamefont {A.}~\bibnamefont
  {L\"auchli}}, \bibinfo {author} {\bibfnamefont {F.}~\bibnamefont {Mila}},\
  and\ \bibinfo {author} {\bibfnamefont {K.}~\bibnamefont {Penc}},\ }\href
  {https://doi.org/10.1103/PhysRevLett.97.087205} {\bibfield  {journal}
  {\bibinfo  {journal} {Phys. Rev. Lett.}\ }\textbf {\bibinfo {volume} {97}},\
  \bibinfo {pages} {087205} (\bibinfo {year} {2006})}\BibitemShut {NoStop}%
\bibitem [{\citenamefont {Nambu}\ \emph {et~al.}(2006)\citenamefont {Nambu},
  \citenamefont {Nakatsuji},\ and\ \citenamefont {Maeno}}]{nambu2006}%
  \BibitemOpen
  \bibfield  {author} {\bibinfo {author} {\bibfnamefont {Y.}~\bibnamefont
  {Nambu}}, \bibinfo {author} {\bibfnamefont {S.}~\bibnamefont {Nakatsuji}},\
  and\ \bibinfo {author} {\bibfnamefont {Y.}~\bibnamefont {Maeno}},\ }\href
  {https://doi.org/10.1143/JPSJ.75.043711} {\bibfield  {journal} {\bibinfo
  {journal} {Journal of the Physical Society of Japan}\ }\textbf {\bibinfo
  {volume} {75}},\ \bibinfo {pages} {043711} (\bibinfo {year}
  {2006})}\BibitemShut {NoStop}%
\bibitem [{\citenamefont {Bhattacharjee}\ \emph {et~al.}(2006)\citenamefont
  {Bhattacharjee}, \citenamefont {Shenoy},\ and\ \citenamefont
  {Senthil}}]{shenoy2006}%
  \BibitemOpen
  \bibfield  {author} {\bibinfo {author} {\bibfnamefont {S.}~\bibnamefont
  {Bhattacharjee}}, \bibinfo {author} {\bibfnamefont {V.~B.}\ \bibnamefont
  {Shenoy}},\ and\ \bibinfo {author} {\bibfnamefont {T.}~\bibnamefont
  {Senthil}},\ }\href {https://doi.org/10.1103/PhysRevB.74.092406} {\bibfield
  {journal} {\bibinfo  {journal} {Phys. Rev. B}\ }\textbf {\bibinfo {volume}
  {74}},\ \bibinfo {pages} {092406} (\bibinfo {year} {2006})}\BibitemShut
  {NoStop}%
\bibitem [{\citenamefont {Cheng}\ \emph {et~al.}(2011)\citenamefont {Cheng},
  \citenamefont {Li}, \citenamefont {Balicas}, \citenamefont {Zhou},
  \citenamefont {Goodenough}, \citenamefont {Xu},\ and\ \citenamefont
  {Zhou}}]{cheng2011high}%
  \BibitemOpen
  \bibfield  {author} {\bibinfo {author} {\bibfnamefont {J.~G.}\ \bibnamefont
  {Cheng}}, \bibinfo {author} {\bibfnamefont {G.}~\bibnamefont {Li}}, \bibinfo
  {author} {\bibfnamefont {L.}~\bibnamefont {Balicas}}, \bibinfo {author}
  {\bibfnamefont {J.~S.}\ \bibnamefont {Zhou}}, \bibinfo {author}
  {\bibfnamefont {J.~B.}\ \bibnamefont {Goodenough}}, \bibinfo {author}
  {\bibfnamefont {C.}~\bibnamefont {Xu}},\ and\ \bibinfo {author}
  {\bibfnamefont {H.~D.}\ \bibnamefont {Zhou}},\ }\href
  {https://doi.org/10.1103/PhysRevLett.107.197204} {\bibfield  {journal}
  {\bibinfo  {journal} {Phys. Rev. Lett.}\ }\textbf {\bibinfo {volume} {107}},\
  \bibinfo {pages} {197204} (\bibinfo {year} {2011})}\BibitemShut {NoStop}%
\bibitem [{\citenamefont {Quilliam}\ \emph {et~al.}(2016)\citenamefont
  {Quilliam}, \citenamefont {Bert}, \citenamefont {Manseau}, \citenamefont
  {Darie}, \citenamefont {Guillot-Deudon}, \citenamefont {Payen}, \citenamefont
  {Baines}, \citenamefont {Amato},\ and\ \citenamefont
  {Mendels}}]{quilliam2016gapless}%
  \BibitemOpen
  \bibfield  {author} {\bibinfo {author} {\bibfnamefont {J.~A.}\ \bibnamefont
  {Quilliam}}, \bibinfo {author} {\bibfnamefont {F.}~\bibnamefont {Bert}},
  \bibinfo {author} {\bibfnamefont {A.}~\bibnamefont {Manseau}}, \bibinfo
  {author} {\bibfnamefont {C.}~\bibnamefont {Darie}}, \bibinfo {author}
  {\bibfnamefont {C.}~\bibnamefont {Guillot-Deudon}}, \bibinfo {author}
  {\bibfnamefont {C.}~\bibnamefont {Payen}}, \bibinfo {author} {\bibfnamefont
  {C.}~\bibnamefont {Baines}}, \bibinfo {author} {\bibfnamefont
  {A.}~\bibnamefont {Amato}},\ and\ \bibinfo {author} {\bibfnamefont
  {P.}~\bibnamefont {Mendels}},\ }\href
  {https://doi.org/10.1103/PhysRevB.93.214432} {\bibfield  {journal} {\bibinfo
  {journal} {Phys. Rev. B}\ }\textbf {\bibinfo {volume} {93}},\ \bibinfo
  {pages} {214432} (\bibinfo {year} {2016})}\BibitemShut {NoStop}%
\bibitem [{\citenamefont {F\aa{}k}\ \emph {et~al.}(2017)\citenamefont
  {F\aa{}k}, \citenamefont {Bieri}, \citenamefont {Can\'evet}, \citenamefont
  {Messio}, \citenamefont {Payen}, \citenamefont {Viaud}, \citenamefont
  {Guillot-Deudon}, \citenamefont {Darie}, \citenamefont {Ollivier},\ and\
  \citenamefont {Mendels}}]{faak2017evidence}%
  \BibitemOpen
  \bibfield  {author} {\bibinfo {author} {\bibfnamefont {B.}~\bibnamefont
  {F\aa{}k}}, \bibinfo {author} {\bibfnamefont {S.}~\bibnamefont {Bieri}},
  \bibinfo {author} {\bibfnamefont {E.}~\bibnamefont {Can\'evet}}, \bibinfo
  {author} {\bibfnamefont {L.}~\bibnamefont {Messio}}, \bibinfo {author}
  {\bibfnamefont {C.}~\bibnamefont {Payen}}, \bibinfo {author} {\bibfnamefont
  {M.}~\bibnamefont {Viaud}}, \bibinfo {author} {\bibfnamefont
  {C.}~\bibnamefont {Guillot-Deudon}}, \bibinfo {author} {\bibfnamefont
  {C.}~\bibnamefont {Darie}}, \bibinfo {author} {\bibfnamefont
  {J.}~\bibnamefont {Ollivier}},\ and\ \bibinfo {author} {\bibfnamefont
  {P.}~\bibnamefont {Mendels}},\ }\href
  {https://doi.org/10.1103/PhysRevB.95.060402} {\bibfield  {journal} {\bibinfo
  {journal} {Phys. Rev. B}\ }\textbf {\bibinfo {volume} {95}},\ \bibinfo
  {pages} {060402} (\bibinfo {year} {2017})}\BibitemShut {NoStop}%
\bibitem [{\citenamefont {Miiller}\ \emph {et~al.}(2011)\citenamefont
  {Miiller}, \citenamefont {Christensen}, \citenamefont {Khan}, \citenamefont
  {Sharma}, \citenamefont {Macquart}, \citenamefont {Avdeev}, \citenamefont
  {McIntyre}, \citenamefont {Piltz},\ and\ \citenamefont
  {Ling}}]{miiller2011yca3}%
  \BibitemOpen
  \bibfield  {author} {\bibinfo {author} {\bibfnamefont {W.}~\bibnamefont
  {Miiller}}, \bibinfo {author} {\bibfnamefont {M.}~\bibnamefont
  {Christensen}}, \bibinfo {author} {\bibfnamefont {A.}~\bibnamefont {Khan}},
  \bibinfo {author} {\bibfnamefont {N.}~\bibnamefont {Sharma}}, \bibinfo
  {author} {\bibfnamefont {R.~B.}\ \bibnamefont {Macquart}}, \bibinfo {author}
  {\bibfnamefont {M.}~\bibnamefont {Avdeev}}, \bibinfo {author} {\bibfnamefont
  {G.~J.}\ \bibnamefont {McIntyre}}, \bibinfo {author} {\bibfnamefont {R.~O.}\
  \bibnamefont {Piltz}},\ and\ \bibinfo {author} {\bibfnamefont {C.~D.}\
  \bibnamefont {Ling}},\ }\href {https://doi.org/10.1021/cm1034003} {\bibfield
  {journal} {\bibinfo  {journal} {Chemistry of Materials}\ }\textbf {\bibinfo
  {volume} {23}},\ \bibinfo {pages} {1315} (\bibinfo {year}
  {2011})}\BibitemShut {NoStop}%
\bibitem [{\citenamefont {Sawaya}\ \emph {et~al.}(2020)\citenamefont {Sawaya},
  \citenamefont {Menke}, \citenamefont {Kyaw}, \citenamefont {Johri},
  \citenamefont {Aspuru-Guzik},\ and\ \citenamefont {Guerreschi}}]{sawaya2020}%
  \BibitemOpen
  \bibfield  {author} {\bibinfo {author} {\bibfnamefont {N.~P.~D.}\
  \bibnamefont {Sawaya}}, \bibinfo {author} {\bibfnamefont {T.}~\bibnamefont
  {Menke}}, \bibinfo {author} {\bibfnamefont {T.~H.}\ \bibnamefont {Kyaw}},
  \bibinfo {author} {\bibfnamefont {S.}~\bibnamefont {Johri}}, \bibinfo
  {author} {\bibfnamefont {A.}~\bibnamefont {Aspuru-Guzik}},\ and\ \bibinfo
  {author} {\bibfnamefont {G.~G.}\ \bibnamefont {Guerreschi}},\ }\href
  {https://doi.org/10.1038/s41534-020-0278-0} {\bibfield  {journal} {\bibinfo
  {journal} {npj Quantum Inf.}\ }\textbf {\bibinfo {volume} {6}},\ \bibinfo
  {pages} {49} (\bibinfo {year} {2020})}\BibitemShut {NoStop}%
\bibitem [{\citenamefont {Di~Matteo}\ \emph {et~al.}(2021)\citenamefont
  {Di~Matteo}, \citenamefont {McCoy}, \citenamefont {Gysbers}, \citenamefont
  {Miyagi}, \citenamefont {Woloshyn},\ and\ \citenamefont
  {Navr\'atil}}]{matteo2021}%
  \BibitemOpen
  \bibfield  {author} {\bibinfo {author} {\bibfnamefont {O.}~\bibnamefont
  {Di~Matteo}}, \bibinfo {author} {\bibfnamefont {A.}~\bibnamefont {McCoy}},
  \bibinfo {author} {\bibfnamefont {P.}~\bibnamefont {Gysbers}}, \bibinfo
  {author} {\bibfnamefont {T.}~\bibnamefont {Miyagi}}, \bibinfo {author}
  {\bibfnamefont {R.~M.}\ \bibnamefont {Woloshyn}},\ and\ \bibinfo {author}
  {\bibfnamefont {P.}~\bibnamefont {Navr\'atil}},\ }\href
  {https://doi.org/10.1103/PhysRevA.103.042405} {\bibfield  {journal} {\bibinfo
   {journal} {Phys. Rev. A}\ }\textbf {\bibinfo {volume} {103}},\ \bibinfo
  {pages} {042405} (\bibinfo {year} {2021})}\BibitemShut {NoStop}%
\bibitem [{\citenamefont {McClean}\ \emph {et~al.}(2016)\citenamefont
  {McClean}, \citenamefont {Romero}, \citenamefont {Babbush},\ and\
  \citenamefont {Aspuru-Guzik}}]{vqe_theory}%
  \BibitemOpen
  \bibfield  {author} {\bibinfo {author} {\bibfnamefont {J.~R.}\ \bibnamefont
  {McClean}}, \bibinfo {author} {\bibfnamefont {J.}~\bibnamefont {Romero}},
  \bibinfo {author} {\bibfnamefont {R.}~\bibnamefont {Babbush}},\ and\ \bibinfo
  {author} {\bibfnamefont {A.}~\bibnamefont {Aspuru-Guzik}},\ }\href
  {https://doi.org/10.1088/1367-2630/18/2/023023} {\bibfield  {journal}
  {\bibinfo  {journal} {New J. Phys.}\ }\textbf {\bibinfo {volume} {18}},\
  \bibinfo {pages} {023023} (\bibinfo {year} {2016})}\BibitemShut {NoStop}%
\bibitem [{\citenamefont {Wecker}\ \emph {et~al.}(2015)\citenamefont {Wecker},
  \citenamefont {Hastings},\ and\ \citenamefont
  {Troyer}}]{wecker2015_trotterizedsp}%
  \BibitemOpen
  \bibfield  {author} {\bibinfo {author} {\bibfnamefont {D.}~\bibnamefont
  {Wecker}}, \bibinfo {author} {\bibfnamefont {M.~B.}\ \bibnamefont
  {Hastings}},\ and\ \bibinfo {author} {\bibfnamefont {M.}~\bibnamefont
  {Troyer}},\ }\href {https://doi.org/10.1103/PhysRevA.92.042303} {\bibfield
  {journal} {\bibinfo  {journal} {Phys. Rev. A}\ }\textbf {\bibinfo {volume}
  {92}},\ \bibinfo {pages} {042303} (\bibinfo {year} {2015})}\BibitemShut
  {NoStop}%
\bibitem [{\citenamefont {Ho}\ and\ \citenamefont
  {Hsieh}(2019)}]{ho2019efficient}%
  \BibitemOpen
  \bibfield  {author} {\bibinfo {author} {\bibfnamefont {W.~W.}\ \bibnamefont
  {Ho}}\ and\ \bibinfo {author} {\bibfnamefont {T.~H.}\ \bibnamefont {Hsieh}},\
  }\href {https://doi.org/10.21468/SciPostPhys.6.3.029} {\bibfield  {journal}
  {\bibinfo  {journal} {SciPost Phys.}\ }\textbf {\bibinfo {volume} {6}},\
  \bibinfo {pages} {29} (\bibinfo {year} {2019})}\BibitemShut {NoStop}%
\bibitem [{\citenamefont {Grimsley}\ \emph {et~al.}(2019)\citenamefont
  {Grimsley}, \citenamefont {Economou}, \citenamefont {Barnes},\ and\
  \citenamefont {Mayhall}}]{grimsleyAdaptiveVariationalAlgorithm2019}%
  \BibitemOpen
  \bibfield  {author} {\bibinfo {author} {\bibfnamefont {H.~R.}\ \bibnamefont
  {Grimsley}}, \bibinfo {author} {\bibfnamefont {S.~E.}\ \bibnamefont
  {Economou}}, \bibinfo {author} {\bibfnamefont {E.}~\bibnamefont {Barnes}},\
  and\ \bibinfo {author} {\bibfnamefont {N.~J.}\ \bibnamefont {Mayhall}},\
  }\href {https://doi.org/10.1038/s41467-019-10988-2} {\bibfield  {journal}
  {\bibinfo  {journal} {Nat. Commun.}\ }\textbf {\bibinfo {volume} {10}},\
  \bibinfo {pages} {3007} (\bibinfo {year} {2019})}\BibitemShut {NoStop}%
\bibitem [{\citenamefont {Tang}\ \emph {et~al.}(2021)\citenamefont {Tang},
  \citenamefont {Shkolnikov}, \citenamefont {Barron}, \citenamefont {Grimsley},
  \citenamefont {Mayhall}, \citenamefont {Barnes},\ and\ \citenamefont
  {Economou}}]{MayhallQubitAVQE}%
  \BibitemOpen
  \bibfield  {author} {\bibinfo {author} {\bibfnamefont {H.~L.}\ \bibnamefont
  {Tang}}, \bibinfo {author} {\bibfnamefont {V.}~\bibnamefont {Shkolnikov}},
  \bibinfo {author} {\bibfnamefont {G.~S.}\ \bibnamefont {Barron}}, \bibinfo
  {author} {\bibfnamefont {H.~R.}\ \bibnamefont {Grimsley}}, \bibinfo {author}
  {\bibfnamefont {N.~J.}\ \bibnamefont {Mayhall}}, \bibinfo {author}
  {\bibfnamefont {E.}~\bibnamefont {Barnes}},\ and\ \bibinfo {author}
  {\bibfnamefont {S.~E.}\ \bibnamefont {Economou}},\ }\href
  {https://doi.org/10.1103/PRXQuantum.2.020310} {\bibfield  {journal} {\bibinfo
   {journal} {PRX Quantum}\ }\textbf {\bibinfo {volume} {2}},\ \bibinfo {pages}
  {020310} (\bibinfo {year} {2021})}\BibitemShut {NoStop}%
\bibitem [{\citenamefont {Wiersema}\ \emph {et~al.}(2020)\citenamefont
  {Wiersema}, \citenamefont {Zhou}, \citenamefont {de~Sereville}, \citenamefont
  {Carrasquilla}, \citenamefont {Kim},\ and\ \citenamefont
  {Yuen}}]{wiersema2020exploring}%
  \BibitemOpen
  \bibfield  {author} {\bibinfo {author} {\bibfnamefont {R.}~\bibnamefont
  {Wiersema}}, \bibinfo {author} {\bibfnamefont {C.}~\bibnamefont {Zhou}},
  \bibinfo {author} {\bibfnamefont {Y.}~\bibnamefont {de~Sereville}}, \bibinfo
  {author} {\bibfnamefont {J.~F.}\ \bibnamefont {Carrasquilla}}, \bibinfo
  {author} {\bibfnamefont {Y.~B.}\ \bibnamefont {Kim}},\ and\ \bibinfo {author}
  {\bibfnamefont {H.}~\bibnamefont {Yuen}},\ }\href
  {https://doi.org/10.1103/PRXQuantum.1.020319} {\bibfield  {journal} {\bibinfo
   {journal} {PRX Quantum}\ }\textbf {\bibinfo {volume} {1}},\ \bibinfo {pages}
  {020319} (\bibinfo {year} {2020})}\BibitemShut {NoStop}%
\bibitem [{\citenamefont {Zhang}\ \emph {et~al.}(2021)\citenamefont {Zhang},
  \citenamefont {Gomes}, \citenamefont {Berthusen}, \citenamefont {Orth},
  \citenamefont {Wang}, \citenamefont {Ho},\ and\ \citenamefont
  {Yao}}]{FengVQE}%
  \BibitemOpen
  \bibfield  {author} {\bibinfo {author} {\bibfnamefont {F.}~\bibnamefont
  {Zhang}}, \bibinfo {author} {\bibfnamefont {N.}~\bibnamefont {Gomes}},
  \bibinfo {author} {\bibfnamefont {N.~F.}\ \bibnamefont {Berthusen}}, \bibinfo
  {author} {\bibfnamefont {P.~P.}\ \bibnamefont {Orth}}, \bibinfo {author}
  {\bibfnamefont {C.-Z.}\ \bibnamefont {Wang}}, \bibinfo {author}
  {\bibfnamefont {K.-M.}\ \bibnamefont {Ho}},\ and\ \bibinfo {author}
  {\bibfnamefont {Y.-X.}\ \bibnamefont {Yao}},\ }\href
  {https://doi.org/10.1103/PhysRevResearch.3.013039} {\bibfield  {journal}
  {\bibinfo  {journal} {Phys. Rev. Res.}\ }\textbf {\bibinfo {volume} {3}},\
  \bibinfo {pages} {013039} (\bibinfo {year} {2021})}\BibitemShut {NoStop}%
\bibitem [{\citenamefont {McClean}\ \emph {et~al.}(2017)\citenamefont
  {McClean}, \citenamefont {Kimchi-Schwartz}, \citenamefont {Carter},\ and\
  \citenamefont {De~Jong}}]{mcclean2017hybrid}%
  \BibitemOpen
  \bibfield  {author} {\bibinfo {author} {\bibfnamefont {J.~R.}\ \bibnamefont
  {McClean}}, \bibinfo {author} {\bibfnamefont {M.~E.}\ \bibnamefont
  {Kimchi-Schwartz}}, \bibinfo {author} {\bibfnamefont {J.}~\bibnamefont
  {Carter}},\ and\ \bibinfo {author} {\bibfnamefont {W.~A.}\ \bibnamefont
  {De~Jong}},\ }\href {https://doi.org/10.1103/PhysRevA.95.042308} {\bibfield
  {journal} {\bibinfo  {journal} {Phys. Rev. A}\ }\textbf {\bibinfo {volume}
  {95}},\ \bibinfo {pages} {042308} (\bibinfo {year} {2017})}\BibitemShut
  {NoStop}%
\bibitem [{\citenamefont {McClean}\ \emph {et~al.}(2020)\citenamefont
  {McClean}, \citenamefont {Jiang}, \citenamefont {Rubin}, \citenamefont
  {Babbush},\ and\ \citenamefont {Neven}}]{mcclean2020}%
  \BibitemOpen
  \bibfield  {author} {\bibinfo {author} {\bibfnamefont {J.~R.}\ \bibnamefont
  {McClean}}, \bibinfo {author} {\bibfnamefont {Z.}~\bibnamefont {Jiang}},
  \bibinfo {author} {\bibfnamefont {N.~C.}\ \bibnamefont {Rubin}}, \bibinfo
  {author} {\bibfnamefont {R.}~\bibnamefont {Babbush}},\ and\ \bibinfo {author}
  {\bibfnamefont {H.}~\bibnamefont {Neven}},\ }\href
  {https://doi.org/10.1038/s41467-020-14341-w} {\bibfield  {journal} {\bibinfo
  {journal} {Nature Communications}\ }\textbf {\bibinfo {volume} {11}},\
  \bibinfo {pages} {636} (\bibinfo {year} {2020})}\BibitemShut {NoStop}%
\bibitem [{\citenamefont {Takeshita}\ \emph {et~al.}(2020)\citenamefont
  {Takeshita}, \citenamefont {Rubin}, \citenamefont {Jiang}, \citenamefont
  {Lee}, \citenamefont {Babbush},\ and\ \citenamefont
  {McClean}}]{takeshita2020}%
  \BibitemOpen
  \bibfield  {author} {\bibinfo {author} {\bibfnamefont {T.}~\bibnamefont
  {Takeshita}}, \bibinfo {author} {\bibfnamefont {N.~C.}\ \bibnamefont
  {Rubin}}, \bibinfo {author} {\bibfnamefont {Z.}~\bibnamefont {Jiang}},
  \bibinfo {author} {\bibfnamefont {E.}~\bibnamefont {Lee}}, \bibinfo {author}
  {\bibfnamefont {R.}~\bibnamefont {Babbush}},\ and\ \bibinfo {author}
  {\bibfnamefont {J.~R.}\ \bibnamefont {McClean}},\ }\href
  {https://doi.org/10.1103/PhysRevX.10.011004} {\bibfield  {journal} {\bibinfo
  {journal} {Phys. Rev. X}\ }\textbf {\bibinfo {volume} {10}},\ \bibinfo
  {pages} {011004} (\bibinfo {year} {2020})}\BibitemShut {NoStop}%
\bibitem [{\citenamefont {Suchsland}\ \emph {et~al.}(2021)\citenamefont
  {Suchsland}, \citenamefont {Tacchino}, \citenamefont {Fischer}, \citenamefont
  {Neupert}, \citenamefont {Barkoutsos},\ and\ \citenamefont
  {Tavernelli}}]{suchsland2021}%
  \BibitemOpen
  \bibfield  {author} {\bibinfo {author} {\bibfnamefont {P.}~\bibnamefont
  {Suchsland}}, \bibinfo {author} {\bibfnamefont {F.}~\bibnamefont {Tacchino}},
  \bibinfo {author} {\bibfnamefont {M.~H.}\ \bibnamefont {Fischer}}, \bibinfo
  {author} {\bibfnamefont {T.}~\bibnamefont {Neupert}}, \bibinfo {author}
  {\bibfnamefont {P.~K.}\ \bibnamefont {Barkoutsos}},\ and\ \bibinfo {author}
  {\bibfnamefont {I.}~\bibnamefont {Tavernelli}},\ }\href
  {https://doi.org/10.22331/q-2021-07-01-492} {\bibfield  {journal} {\bibinfo
  {journal} {{Quantum}}\ }\textbf {\bibinfo {volume} {5}},\ \bibinfo {pages}
  {492} (\bibinfo {year} {2021})}\BibitemShut {NoStop}%
\bibitem [{\citenamefont {Yoshioka}\ \emph {et~al.}(2022)\citenamefont
  {Yoshioka}, \citenamefont {Hakoshima}, \citenamefont {Matsuzaki},
  \citenamefont {Tokunaga}, \citenamefont {Suzuki},\ and\ \citenamefont
  {Endo}}]{yoshioka2022}%
  \BibitemOpen
  \bibfield  {author} {\bibinfo {author} {\bibfnamefont {N.}~\bibnamefont
  {Yoshioka}}, \bibinfo {author} {\bibfnamefont {H.}~\bibnamefont {Hakoshima}},
  \bibinfo {author} {\bibfnamefont {Y.}~\bibnamefont {Matsuzaki}}, \bibinfo
  {author} {\bibfnamefont {Y.}~\bibnamefont {Tokunaga}}, \bibinfo {author}
  {\bibfnamefont {Y.}~\bibnamefont {Suzuki}},\ and\ \bibinfo {author}
  {\bibfnamefont {S.}~\bibnamefont {Endo}},\ }\href
  {https://doi.org/10.1103/PhysRevLett.129.020502} {\bibfield  {journal}
  {\bibinfo  {journal} {Phys. Rev. Lett.}\ }\textbf {\bibinfo {volume} {129}},\
  \bibinfo {pages} {020502} (\bibinfo {year} {2022})}\BibitemShut {NoStop}%
\bibitem [{\citenamefont {Beach}\ \emph {et~al.}(2019)\citenamefont {Beach},
  \citenamefont {Melko}, \citenamefont {Grover},\ and\ \citenamefont
  {Hsieh}}]{beach2019making}%
  \BibitemOpen
  \bibfield  {author} {\bibinfo {author} {\bibfnamefont {M.~J.}\ \bibnamefont
  {Beach}}, \bibinfo {author} {\bibfnamefont {R.~G.}\ \bibnamefont {Melko}},
  \bibinfo {author} {\bibfnamefont {T.}~\bibnamefont {Grover}},\ and\ \bibinfo
  {author} {\bibfnamefont {T.~H.}\ \bibnamefont {Hsieh}},\ }\href
  {https://doi.org/10.1103/PhysRevB.100.094434} {\bibfield  {journal} {\bibinfo
   {journal} {Phys. Rev. B}\ }\textbf {\bibinfo {volume} {100}},\ \bibinfo
  {pages} {094434} (\bibinfo {year} {2019})}\BibitemShut {NoStop}%
\bibitem [{\citenamefont {McArdle}\ \emph {et~al.}(2019)\citenamefont
  {McArdle}, \citenamefont {Jones}, \citenamefont {Endo}, \citenamefont {Li},
  \citenamefont {Benjamin},\ and\ \citenamefont {Yuan}}]{VQITE}%
  \BibitemOpen
  \bibfield  {author} {\bibinfo {author} {\bibfnamefont {S.}~\bibnamefont
  {McArdle}}, \bibinfo {author} {\bibfnamefont {T.}~\bibnamefont {Jones}},
  \bibinfo {author} {\bibfnamefont {S.}~\bibnamefont {Endo}}, \bibinfo {author}
  {\bibfnamefont {Y.}~\bibnamefont {Li}}, \bibinfo {author} {\bibfnamefont
  {S.~C.}\ \bibnamefont {Benjamin}},\ and\ \bibinfo {author} {\bibfnamefont
  {X.}~\bibnamefont {Yuan}},\ }\href
  {https://doi.org/10.1038/s41534-019-0187-2} {\bibfield  {journal} {\bibinfo
  {journal} {npj Quantum Inf.}\ }\textbf {\bibinfo {volume} {5}},\ \bibinfo
  {pages} {75} (\bibinfo {year} {2019})}\BibitemShut {NoStop}%
\bibitem [{\citenamefont {Stokes}\ \emph {et~al.}(2020)\citenamefont {Stokes},
  \citenamefont {Izaac}, \citenamefont {Killoran},\ and\ \citenamefont
  {Carleo}}]{stokes2020quantum}%
  \BibitemOpen
  \bibfield  {author} {\bibinfo {author} {\bibfnamefont {J.}~\bibnamefont
  {Stokes}}, \bibinfo {author} {\bibfnamefont {J.}~\bibnamefont {Izaac}},
  \bibinfo {author} {\bibfnamefont {N.}~\bibnamefont {Killoran}},\ and\
  \bibinfo {author} {\bibfnamefont {G.}~\bibnamefont {Carleo}},\ }\href
  {https://doi.org/10.22331/q-2020-05-25-269} {\bibfield  {journal} {\bibinfo
  {journal} {Quantum}\ }\textbf {\bibinfo {volume} {4}},\ \bibinfo {pages}
  {269} (\bibinfo {year} {2020})}\BibitemShut {NoStop}%
\bibitem [{\citenamefont {Nishi}\ \emph
  {et~al.}(2021{\natexlab{a}})\citenamefont {Nishi}, \citenamefont {Kosugi},\
  and\ \citenamefont {Matsushita}}]{qite_nla}%
  \BibitemOpen
  \bibfield  {author} {\bibinfo {author} {\bibfnamefont {H.}~\bibnamefont
  {Nishi}}, \bibinfo {author} {\bibfnamefont {T.}~\bibnamefont {Kosugi}},\ and\
  \bibinfo {author} {\bibfnamefont {Y.-i.}\ \bibnamefont {Matsushita}},\ }\href
  {https://doi.org/10.1038/s41534-021-00409-y} {\bibfield  {journal} {\bibinfo
  {journal} {npj Quantum Information}\ }\textbf {\bibinfo {volume} {7}},\
  \bibinfo {pages} {85} (\bibinfo {year} {2021}{\natexlab{a}})}\BibitemShut
  {NoStop}%
\bibitem [{\citenamefont {Gomes}\ \emph {et~al.}(2020)\citenamefont {Gomes},
  \citenamefont {Zhang}, \citenamefont {Berthusen}, \citenamefont {Wang},
  \citenamefont {Ho}, \citenamefont {Orth},\ and\ \citenamefont
  {Yao}}]{smqite}%
  \BibitemOpen
  \bibfield  {author} {\bibinfo {author} {\bibfnamefont {N.}~\bibnamefont
  {Gomes}}, \bibinfo {author} {\bibfnamefont {F.}~\bibnamefont {Zhang}},
  \bibinfo {author} {\bibfnamefont {N.~F.}\ \bibnamefont {Berthusen}}, \bibinfo
  {author} {\bibfnamefont {C.-Z.}\ \bibnamefont {Wang}}, \bibinfo {author}
  {\bibfnamefont {K.-M.}\ \bibnamefont {Ho}}, \bibinfo {author} {\bibfnamefont
  {P.~P.}\ \bibnamefont {Orth}},\ and\ \bibinfo {author} {\bibfnamefont
  {Y.-X.}\ \bibnamefont {Yao}},\ }\href
  {https://doi.org/10.1021/acs.jctc.0c00666} {\bibfield  {journal} {\bibinfo
  {journal} {J. Chem. Theory Comput.}\ }\textbf {\bibinfo {volume} {16}},\
  \bibinfo {pages} {6256} (\bibinfo {year} {2020})}\BibitemShut {NoStop}%
\bibitem [{\citenamefont {Yeter-Aydeniz}\ \emph {et~al.}(2020)\citenamefont
  {Yeter-Aydeniz}, \citenamefont {Pooser},\ and\ \citenamefont
  {Siopsis}}]{QITE_h2}%
  \BibitemOpen
  \bibfield  {author} {\bibinfo {author} {\bibfnamefont {K.}~\bibnamefont
  {Yeter-Aydeniz}}, \bibinfo {author} {\bibfnamefont {R.~C.}\ \bibnamefont
  {Pooser}},\ and\ \bibinfo {author} {\bibfnamefont {G.}~\bibnamefont
  {Siopsis}},\ }\href {https://doi.org/10.1038/s41534-020-00290-1} {\bibfield
  {journal} {\bibinfo  {journal} {npj Quantum Inf.}\ }\textbf {\bibinfo
  {volume} {6}},\ \bibinfo {pages} {1} (\bibinfo {year} {2020})}\BibitemShut
  {NoStop}%
\bibitem [{\citenamefont {Sun}\ \emph {et~al.}(2021)\citenamefont {Sun},
  \citenamefont {Motta}, \citenamefont {Tazhigulov}, \citenamefont {Tan},
  \citenamefont {Chan},\ and\ \citenamefont {Minnich}}]{Sun2021QuantumCO}%
  \BibitemOpen
  \bibfield  {author} {\bibinfo {author} {\bibfnamefont {S.-N.}\ \bibnamefont
  {Sun}}, \bibinfo {author} {\bibfnamefont {M.}~\bibnamefont {Motta}}, \bibinfo
  {author} {\bibfnamefont {R.~N.}\ \bibnamefont {Tazhigulov}}, \bibinfo
  {author} {\bibfnamefont {A.~T.}\ \bibnamefont {Tan}}, \bibinfo {author}
  {\bibfnamefont {G.~K.-L.}\ \bibnamefont {Chan}},\ and\ \bibinfo {author}
  {\bibfnamefont {A.~J.}\ \bibnamefont {Minnich}},\ }\href
  {https://doi.org/10.1103/PRXQuantum.2.010317} {\bibfield  {journal} {\bibinfo
   {journal} {PRX Quantum}\ }\textbf {\bibinfo {volume} {2}},\ \bibinfo {pages}
  {010317} (\bibinfo {year} {2021})}\BibitemShut {NoStop}%
\bibitem [{\citenamefont {Nishi}\ \emph
  {et~al.}(2021{\natexlab{b}})\citenamefont {Nishi}, \citenamefont {Kosugi},\
  and\ \citenamefont {ichiro Matsushita}}]{Nishi2021ImplementationOQ}%
  \BibitemOpen
  \bibfield  {author} {\bibinfo {author} {\bibfnamefont {H.}~\bibnamefont
  {Nishi}}, \bibinfo {author} {\bibfnamefont {T.}~\bibnamefont {Kosugi}},\ and\
  \bibinfo {author} {\bibfnamefont {Y.}~\bibnamefont {ichiro Matsushita}},\
  }\href {https://doi.org/10.1038/s41534-021-00409-y} {\bibfield  {journal}
  {\bibinfo  {journal} {npj Quantum Inf.}\ }\textbf {\bibinfo {volume} {7}},\
  \bibinfo {pages} {1} (\bibinfo {year} {2021}{\natexlab{b}})}\BibitemShut
  {NoStop}%
\bibitem [{\citenamefont {Gomes}\ \emph {et~al.}(2021)\citenamefont {Gomes},
  \citenamefont {Mukherjee}, \citenamefont {Zhang}, \citenamefont {Iadecola},
  \citenamefont {Wang}, \citenamefont {Ho}, \citenamefont {Orth},\ and\
  \citenamefont {Yao}}]{AVQITE}%
  \BibitemOpen
  \bibfield  {author} {\bibinfo {author} {\bibfnamefont {N.}~\bibnamefont
  {Gomes}}, \bibinfo {author} {\bibfnamefont {A.}~\bibnamefont {Mukherjee}},
  \bibinfo {author} {\bibfnamefont {F.}~\bibnamefont {Zhang}}, \bibinfo
  {author} {\bibfnamefont {T.}~\bibnamefont {Iadecola}}, \bibinfo {author}
  {\bibfnamefont {C.-Z.}\ \bibnamefont {Wang}}, \bibinfo {author}
  {\bibfnamefont {K.-M.}\ \bibnamefont {Ho}}, \bibinfo {author} {\bibfnamefont
  {P.~P.}\ \bibnamefont {Orth}},\ and\ \bibinfo {author} {\bibfnamefont
  {Y.-X.}\ \bibnamefont {Yao}},\ }\href
  {https://doi.org/10.1002/qute.202100114} {\bibfield  {journal} {\bibinfo
  {journal} {Adv. Quantum Technol.}\ }\textbf {\bibinfo {volume} {4}},\
  \bibinfo {pages} {2100114} (\bibinfo {year} {2021})}\BibitemShut {NoStop}%
\bibitem [{\citenamefont {Mukherjee}\ \emph {et~al.}(2023)\citenamefont
  {Mukherjee}, \citenamefont {Berthusen}, \citenamefont {Getelina},
  \citenamefont {Orth},\ and\ \citenamefont {Yao}}]{mukherjee2023comparative}%
  \BibitemOpen
  \bibfield  {author} {\bibinfo {author} {\bibfnamefont {A.}~\bibnamefont
  {Mukherjee}}, \bibinfo {author} {\bibfnamefont {N.~F.}\ \bibnamefont
  {Berthusen}}, \bibinfo {author} {\bibfnamefont {J.~C.}\ \bibnamefont
  {Getelina}}, \bibinfo {author} {\bibfnamefont {P.~P.}\ \bibnamefont {Orth}},\
  and\ \bibinfo {author} {\bibfnamefont {Y.-X.}\ \bibnamefont {Yao}},\ }\href
  {https://doi.org/10.1038/s42005-022-01089-6} {\bibfield  {journal} {\bibinfo
  {journal} {Commun. Phys.}\ }\textbf {\bibinfo {volume} {6}},\ \bibinfo
  {pages} {4} (\bibinfo {year} {2023})}\BibitemShut {NoStop}%
\bibitem [{\citenamefont {Li}\ and\ \citenamefont
  {Benjamin}(2017)}]{VDynamics_Li}%
  \BibitemOpen
  \bibfield  {author} {\bibinfo {author} {\bibfnamefont {Y.}~\bibnamefont
  {Li}}\ and\ \bibinfo {author} {\bibfnamefont {S.~C.}\ \bibnamefont
  {Benjamin}},\ }\href {https://doi.org/10.1103/PhysRevX.7.021050} {\bibfield
  {journal} {\bibinfo  {journal} {Phys. Rev. X}\ }\textbf {\bibinfo {volume}
  {7}},\ \bibinfo {pages} {021050} (\bibinfo {year} {2017})}\BibitemShut
  {NoStop}%
\bibitem [{\citenamefont {Blume}(1966)}]{blume1966}%
  \BibitemOpen
  \bibfield  {author} {\bibinfo {author} {\bibfnamefont {M.}~\bibnamefont
  {Blume}},\ }\href {https://doi.org/10.1103/PhysRev.141.517} {\bibfield
  {journal} {\bibinfo  {journal} {Phys. Rev.}\ }\textbf {\bibinfo {volume}
  {141}},\ \bibinfo {pages} {517} (\bibinfo {year} {1966})}\BibitemShut
  {NoStop}%
\bibitem [{\citenamefont {Capel}(1966)}]{capel1966}%
  \BibitemOpen
  \bibfield  {author} {\bibinfo {author} {\bibfnamefont {H.}~\bibnamefont
  {Capel}},\ }\href {https://doi.org/10.1016/0031-8914(66)90027-9} {\bibfield
  {journal} {\bibinfo  {journal} {Physica}\ }\textbf {\bibinfo {volume} {32}},\
  \bibinfo {pages} {966} (\bibinfo {year} {1966})}\BibitemShut {NoStop}%
\bibitem [{\citenamefont {Blume}\ \emph {et~al.}(1971)\citenamefont {Blume},
  \citenamefont {Emery},\ and\ \citenamefont {Griffiths}}]{blume1971}%
  \BibitemOpen
  \bibfield  {author} {\bibinfo {author} {\bibfnamefont {M.}~\bibnamefont
  {Blume}}, \bibinfo {author} {\bibfnamefont {V.~J.}\ \bibnamefont {Emery}},\
  and\ \bibinfo {author} {\bibfnamefont {R.~B.}\ \bibnamefont {Griffiths}},\
  }\href {https://doi.org/10.1103/PhysRevA.4.1071} {\bibfield  {journal}
  {\bibinfo  {journal} {Phys. Rev. A}\ }\textbf {\bibinfo {volume} {4}},\
  \bibinfo {pages} {1071} (\bibinfo {year} {1971})}\BibitemShut {NoStop}%
\bibitem [{\citenamefont {Kaufman}\ \emph {et~al.}(1981)\citenamefont
  {Kaufman}, \citenamefont {Griffiths}, \citenamefont {Yeomans},\ and\
  \citenamefont {Fisher}}]{kaufman1981}%
  \BibitemOpen
  \bibfield  {author} {\bibinfo {author} {\bibfnamefont {M.}~\bibnamefont
  {Kaufman}}, \bibinfo {author} {\bibfnamefont {R.~B.}\ \bibnamefont
  {Griffiths}}, \bibinfo {author} {\bibfnamefont {J.~M.}\ \bibnamefont
  {Yeomans}},\ and\ \bibinfo {author} {\bibfnamefont {M.~E.}\ \bibnamefont
  {Fisher}},\ }\href {https://doi.org/10.1103/PhysRevB.23.3448} {\bibfield
  {journal} {\bibinfo  {journal} {Phys. Rev. B}\ }\textbf {\bibinfo {volume}
  {23}},\ \bibinfo {pages} {3448} (\bibinfo {year} {1981})}\BibitemShut
  {NoStop}%
\bibitem [{\citenamefont {Berker}\ and\ \citenamefont
  {Wortis}(1976)}]{berker1976}%
  \BibitemOpen
  \bibfield  {author} {\bibinfo {author} {\bibfnamefont {A.~N.}\ \bibnamefont
  {Berker}}\ and\ \bibinfo {author} {\bibfnamefont {M.}~\bibnamefont
  {Wortis}},\ }\href {https://doi.org/10.1103/PhysRevB.14.4946} {\bibfield
  {journal} {\bibinfo  {journal} {Phys. Rev. B}\ }\textbf {\bibinfo {volume}
  {14}},\ \bibinfo {pages} {4946} (\bibinfo {year} {1976})}\BibitemShut
  {NoStop}%
\bibitem [{\citenamefont {Burkhardt}(1976)}]{burkhardt1976}%
  \BibitemOpen
  \bibfield  {author} {\bibinfo {author} {\bibfnamefont {T.~W.}\ \bibnamefont
  {Burkhardt}},\ }\href {https://doi.org/10.1103/PhysRevB.14.1196} {\bibfield
  {journal} {\bibinfo  {journal} {Phys. Rev. B}\ }\textbf {\bibinfo {volume}
  {14}},\ \bibinfo {pages} {1196} (\bibinfo {year} {1976})}\BibitemShut
  {NoStop}%
\bibitem [{\citenamefont {Alcaraz}\ \emph {et~al.}(1985)\citenamefont
  {Alcaraz}, \citenamefont {Drugowich~de Fel\'{\i}cio}, \citenamefont
  {K\"oberle},\ and\ \citenamefont {Stilck}}]{alcaraz1985}%
  \BibitemOpen
  \bibfield  {author} {\bibinfo {author} {\bibfnamefont {F.~C.}\ \bibnamefont
  {Alcaraz}}, \bibinfo {author} {\bibfnamefont {J.~R.}\ \bibnamefont
  {Drugowich~de Fel\'{\i}cio}}, \bibinfo {author} {\bibfnamefont
  {R.}~\bibnamefont {K\"oberle}},\ and\ \bibinfo {author} {\bibfnamefont
  {J.~F.}\ \bibnamefont {Stilck}},\ }\href
  {https://doi.org/10.1103/PhysRevB.32.7469} {\bibfield  {journal} {\bibinfo
  {journal} {Phys. Rev. B}\ }\textbf {\bibinfo {volume} {32}},\ \bibinfo
  {pages} {7469} (\bibinfo {year} {1985})}\BibitemShut {NoStop}%
\bibitem [{\citenamefont {Qiu}(1986)}]{qiu1986}%
  \BibitemOpen
  \bibfield  {author} {\bibinfo {author} {\bibfnamefont {Z.}~\bibnamefont
  {Qiu}},\ }\href {https://doi.org/10.1016/0550-3213(86)90553-5} {\bibfield
  {journal} {\bibinfo  {journal} {Nucl. Phys. B.}\ }\textbf {\bibinfo {volume}
  {270}},\ \bibinfo {pages} {205} (\bibinfo {year} {1986})}\BibitemShut
  {NoStop}%
\bibitem [{\citenamefont {Balbao}\ and\ \citenamefont
  {de~Felicio}(1987)}]{balbao1987}%
  \BibitemOpen
  \bibfield  {author} {\bibinfo {author} {\bibfnamefont {D.~B.}\ \bibnamefont
  {Balbao}}\ and\ \bibinfo {author} {\bibfnamefont {J.~R.~D.}\ \bibnamefont
  {de~Felicio}},\ }\href {https://doi.org/10.1088/0305-4470/20/4/005}
  {\bibfield  {journal} {\bibinfo  {journal} {Journal of Physics A:
  Mathematical and General}\ }\textbf {\bibinfo {volume} {20}},\ \bibinfo
  {pages} {L207} (\bibinfo {year} {1987})}\BibitemShut {NoStop}%
\bibitem [{\citenamefont {Chen}\ \emph {et~al.}(2003)\citenamefont {Chen},
  \citenamefont {Hida},\ and\ \citenamefont {Sanctuary}}]{chen2003}%
  \BibitemOpen
  \bibfield  {author} {\bibinfo {author} {\bibfnamefont {W.}~\bibnamefont
  {Chen}}, \bibinfo {author} {\bibfnamefont {K.}~\bibnamefont {Hida}},\ and\
  \bibinfo {author} {\bibfnamefont {B.~C.}\ \bibnamefont {Sanctuary}},\ }\href
  {https://doi.org/10.1103/PhysRevB.67.104401} {\bibfield  {journal} {\bibinfo
  {journal} {Phys. Rev. B}\ }\textbf {\bibinfo {volume} {67}},\ \bibinfo
  {pages} {104401} (\bibinfo {year} {2003})}\BibitemShut {NoStop}%
\bibitem [{\citenamefont {Wang}\ \emph {et~al.}(2020)\citenamefont {Wang},
  \citenamefont {Hu}, \citenamefont {Sanders},\ and\ \citenamefont
  {Kais}}]{wang2020}%
  \BibitemOpen
  \bibfield  {author} {\bibinfo {author} {\bibfnamefont {Y.}~\bibnamefont
  {Wang}}, \bibinfo {author} {\bibfnamefont {Z.}~\bibnamefont {Hu}}, \bibinfo
  {author} {\bibfnamefont {B.~C.}\ \bibnamefont {Sanders}},\ and\ \bibinfo
  {author} {\bibfnamefont {S.}~\bibnamefont {Kais}},\ }\href
  {https://doi.org/10.3389/fphy.2020.589504} {\bibfield  {journal} {\bibinfo
  {journal} {Frontiers in Physics}\ }\textbf {\bibinfo {volume} {8}},\ \bibinfo
  {pages} {1} (\bibinfo {year} {2020})}\BibitemShut {NoStop}%
\bibitem [{\citenamefont {{Accessing Higher Energy States with Qiskit
  Pulse}}()}]{qiskit_qudit}%
  \BibitemOpen
  \bibfield  {author} {\bibinfo {author} {\bibnamefont {{Accessing Higher
  Energy States with Qiskit Pulse}}},\ }\href
  {https://learn.qiskit.org/course/quantum-hardware-pulses/accessing-higher-energy-states-with-qiskit-pulse}
  {\bibinfo  {journal}
  {https://learn.qiskit.org/course/quantum-hardware-pulses}\ }\BibitemShut
  {NoStop}%
\bibitem [{\citenamefont {Yuan}\ \emph {et~al.}(2019)\citenamefont {Yuan},
  \citenamefont {Endo}, \citenamefont {Zhao}, \citenamefont {Li},\ and\
  \citenamefont {Benjamin}}]{theory_vqs}%
  \BibitemOpen
\bibfield  {journal} {  }\bibfield  {author} {\bibinfo {author} {\bibfnamefont
  {X.}~\bibnamefont {Yuan}}, \bibinfo {author} {\bibfnamefont {S.}~\bibnamefont
  {Endo}}, \bibinfo {author} {\bibfnamefont {Q.}~\bibnamefont {Zhao}}, \bibinfo
  {author} {\bibfnamefont {Y.}~\bibnamefont {Li}},\ and\ \bibinfo {author}
  {\bibfnamefont {S.~C.}\ \bibnamefont {Benjamin}},\ }\href
  {https://doi.org/10.22331/q-2019-10-07-191} {\bibfield  {journal} {\bibinfo
  {journal} {Quantum}\ }\textbf {\bibinfo {volume} {3}},\ \bibinfo {pages}
  {191} (\bibinfo {year} {2019})}\BibitemShut {NoStop}%
\end{thebibliography}%

\appendix
\begin{widetext}
\section{\label{sec:ed}Pinpointing continuous phase transitions with exact diagonalization}
\begin{figure*}[b!]
\begin{center}
    \includegraphics[width=\linewidth]{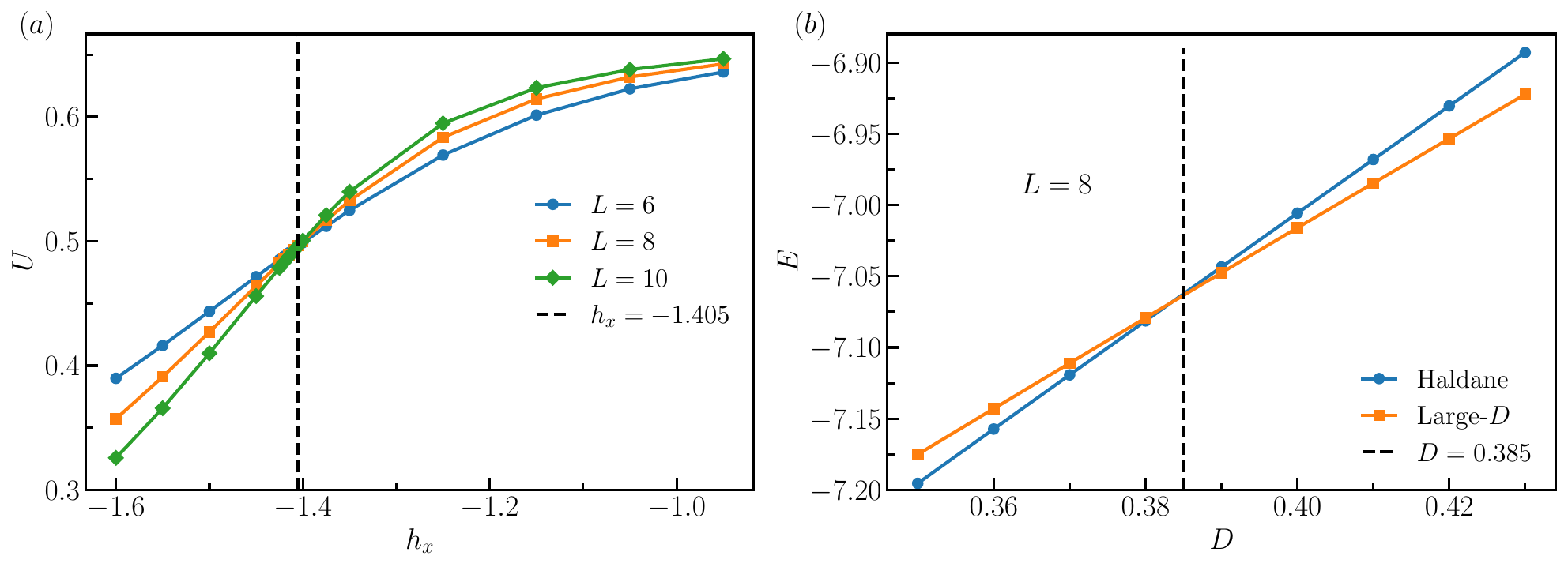}\caption{\label{fig:data-ED}
	(a) Binder cumulant $U$ as a function of the external magnetic field $h_x$ for the Blume-Capel model
	with various system sizes $L$. (b) Energy $E$ of the lowest-lying states in the Haldane and
	large-$D$ phases vs. the single-ion anisotropy $D$ for a XXZ spin-1 chain of size $L=8$.}
\par\end{center}
\end{figure*}
In this appendix we explain the procedure adopted to determine the model parameter sets that are located close to the quantum phase transition (in the thermodynamic limit), for which we employ exact diagonalization.

As mentioned in the main text, for the Blume-Capel model there is a continuous phase transition line separating
the ordered and disordered phases (see Fig.~\ref{fig:phase_diag}). In cases where one is limited to relatively
small system sizes, a good approach to identify a critical point
is to measure the so-called Binder cumulant $U$, which is defined as
\begin{equation}
	U = 1 - \frac{\left\langle m^4 \right\rangle}{3\left\langle m^2 \right\rangle^2},
\end{equation}
where $m = \left( \sum_i^L S_i^z \right) / L$ is the magnetization, and $L$ is the spin
chain length. The Binder cumulant
has the crucial property that its scaling dimension vanishes at criticality. Thus, 
one can precisely pinpoint phase transitions by observing the crossing of $U$ curves
for different system sizes, plotted as a function of a
phase-transition driving parameter.
In Fig.~\ref{fig:data-ED}\hyperlink{fig:data-ED}{(a)} we show the Binder cumulant for several spin
chain lengths as a function of the transverse field strength $h_x$. The dashed line
denotes the point in which the curves cross, which is approximately $h_x=1.405$.

Conversely, for the XXZ model with single-ion anisotropy, we employ the method described
in Ref.~\cite{chen2003} to estimate a critical point for the transition between the large-$D$
and the Haldane phase. The approach consists of measuring the energy of the lowest state in
two different sectors, which are related to the sign of the eigenvalues of the space inversion
and spin reversal operators. Due to the existence of edge states in the Haldane phase, a
system with twisted boundary conditions has an eigenvalue equal to $-1$ for the two aforementioned
inversion operators, whereas in the large-$D$ phase this eigenvalue is equal to $+1$ as there
are no symmetry protected edge states. Therefore, whenever the state in the sector $-1$ has
smaller energy, the system finds itself in the Haldane phase, and vice-versa.
Fig.~\ref{fig:data-ED}\hyperlink{fig:data-ED}{(b)} displays the lowest-energy curves pertaining to these
two sectors as a function of the strength of the single-ion anisotropy $D$. The crossing
point, which reads $D=0.385$ as indicated by the dashed line, tells precisely the location of the phase transition.
\end{widetext}

\end{document}